\documentclass[sigconf]{acmart}

\usepackage{soul}
\usepackage{url}
\usepackage[utf8]{inputenc}
\usepackage{graphicx}
\usepackage{amsmath}
\usepackage{amsthm}
\usepackage{booktabs}
\usepackage{algorithm}
\usepackage{algorithmic}
\usepackage{array}
\usepackage{float}
\usepackage{caption}
\usepackage{subcaption}
\usepackage{xcolor}
\usepackage{enumitem}

\newcolumntype{P}[1]{>{\centering\arraybackslash}p{#1}}
\newcolumntype{M}[1]{>{\centering\arraybackslash}m{#1}}

{

\title{Characterizing Human Mobility Patterns During COVID-19 using Cellular Network Data}

\author[Ayan et al.]{Necati A. Ayan${^1}$, Nilson L. Damasceno${^2}$, Sushil Chaskar${^1}$, Peron R. de Sousa${^2}$, Arti Ramesh${^1}$, Anand Seetharam${^1}$, and  Antonio A. de A. Rocha${^2}$\\
\textnormal{SUNY Binghamton}{$^1$} \hspace{0.1cm} \textnormal{Institute of Computing, Fluminense Federal University}{$^2$}\\
\textnormal{\{nayan1, schaska1, artir, aseethar\}@binghamton.edu \{nilsonld, peron\_rezende, arocha\}@id.uff.br}}}

\begin{document}




\begin{abstract}
In this paper, our goal is to analyze and compare cellular network usage data from  pre-lockdown, during lockdown, and post-lockdown phases surrounding the COVID-19 pandemic to understand  and model human mobility patterns during the pandemic, and evaluate the effect of lockdowns on mobility. To this end, we collaborate with one of the main cellular network providers in Brazil, and collect and analyze cellular network connections from 1400 antennas for all users in the city of Rio de Janeiro and its suburbs from  March 1, 2020 to July 1, 2020. Our analysis reveals that the total number of cellular connections decreases to 78\% during the lockdown phase and then increases to 85\% of the pre-COVID era as the lockdown eases. We  observe that as more people work remotely, there is a shift in the antennas incurring top 10\% of the total traffic, with the number of connections made to antennas in downtown Rio reducing drastically and antennas at other locations taking their place. We also observe that while nearly 40-45\% users connected to only 1 antenna each day during the lockdown phase indicating no mobility, there are around 4\% users (i.e., 80K users) who connected to more than 10 antennas, indicating very high mobility. We also observe that the amount of mobility increases towards the end of the lockdown period even before the lockdown eases and the upward trend continues in the post-lockdown period. Finally, we design an interactive tool that showcases mobility patterns in different granularities that can potentially help people and government officials understand the mobility of individuals and the number of COVID cases in a particular neighborhood. Our analysis, inferences, and interactive showcasing of mobility patterns based on large-scale data can be extrapolated to other cities of the world and has the potential to help in designing more effective pandemic management measures in the future.

\end{abstract}

%
%

%
%


\maketitle


\section{Introduction}
\label{sec:intro}

COVID-19 is a global pandemic that has infected human beings in all countries of the world. Lack of physical distancing, isolation, mask-wearing,  and effective contact tracing are some of the key factors that have contributed to the spread of COVID-19 and transformed it into a global pandemic.  To mitigate the spread of the disease, most countries around the world have implemented varying levels of lockdown.  Though lockdowns have been effective in decreasing the rate of spread \cite{atalan2020lockdown}, they have not been able to curb the disease and infections continue to soar in countries around the world. Therefore, there is an urgent need to understand the effects of lockdown on mobility patterns, so they can be effectively integrated into government policies to manage the COVID-19 pandemic.

In this paper, our goal is to analyze and compare cellular network usage data (comprising of phone calls, 3G/4G data connections, and text messages) from pre-lockdown, during lockdown, and post-lockdown phases of the COVID-19 pandemic to  understand  and model human mobility patterns, and evaluate the effect of lockdown on mobility. To this end, we collaborate with one of the main cellular network providers in Brazil, TIM Brazil, and conduct a large scale study by collecting and analyzing anonymized cellular network connections from all users in the city of Rio de Janeiro, the second most populous city in Brazil, and its suburbs. The data consists of individual connections made by users to approximately 1400 cellular antennas in and around the city of Rio de Janeiro and its suburbs during each 5-minute interval from March 1, 2020 to July 1, 2020. There are approximately 120 million connections for each day made by approximately 2 million users per day during this time period in our dataset, amounting to a total of approximately 10 billion connection logs. As Brazil enforced strict lockdown measures between the third week of March and end of May, the data and the ensuing analysis provides valuable insight into human behavior and mobility in the pre-lockdown, during lockdown, and post-lockdown time periods, making it a comprehensive study of mobility that can offer valuable perspective for effective lockdown and pandemic management.  Our main contributions are summarized below.


\noindent \textbf{Connectivity and User Mobility Analysis:} 
Our data analysis reveals some interesting trends. We observe that the total number of cellular connections decreases to 78\% during lockdown  and then increases again to 85\% of the pre-lockdown values during the post-lockdown period. The  number of distinct users using cellular network connections also increases in the post-lockdown period as more people venture outside and thus need to use the cellular network. To investigate the impact of lockdown on mobility, we investigate the top 10\% of antennas (i.e., 140) that carry the highest amount of traffic. We observe that new antennas emerge in the top 10\% both in the lockdown and post-lockdown phases that replace some of the antennas in the top 10\% in the pre-lockdown phase. A closer look reveals that  antennas that serve downtown Rio de Janeiro as well as those that serve the commercial hubs of some of the main districts of the city no longer feature in the top 10\% of antennas during lockdown. We also conduct user mobility analysis and observe that while approximately 35--40\% users exhibit no mobility (i.e., connect to a single antenna per day) during the lockdown and post-lockdown periods, approximately 4\% of users (i.e., 80K users) exhibit high mobility (i.e., connect to more than 10 antenna per day). This high mobility is interesting as it is likely to be demonstrated by essential workers and those flouting lockdown measures. 

\noindent \textbf{Graph-based User Mobility Analysis:} 
We next conduct a graph-based analysis  to better understand and model the mobility patterns of people during COVID-19. To this end, for each day, we construct a graph where the antennas correspond to the vertices and the movement of users between antennas corresponds to the weight of that particular edge. We determine the total in-degree of the nodes of the graph to quantify the total number of mobility events and observe that user mobility starts increasing around 3 weeks before the end of lockdown, with the trend continuing into the post-lockdown period. We also investigate the impact of the day of the week on mobility. We observe that weekdays and Saturday have similar levels of mobility during and after lockdown while Sunday has the least mobility. This difference in mobility between the other days of the week and Sunday is interesting because it suggests that many people do not have the opportunity to work from home even during the lockdown period.

We next construct heatmaps for mobility by grouping the antennas into main municipal regions of the city to identify: {\it i)} the geographical regions that have the highest mobility, and {\it ii)} to investigate the change in the mobility of particular antennas during and post lockdown. Interestingly, we detect a connection between mobility patterns observed at antennas and the social progress index (SPI) of the region in which they are located. We observe that antennas in regions having a low SPI often exhibit higher mobility when compared to the ones in regions having a higher SPI. 

\noindent \textbf{COVID-19 Borescope:} 
Finally, we design a visual/interactive tool, COVID-19 Borescope, which helps people and government administrators analyze the mobility of individuals as well as correlate it with the number of COVID-19 cases in the city. To deal with the massive amount of data, we use an optimized version of the nanocubes data structure to make the tool scalable and highly interactive. The interactive web interface provides multiple functionalities including selecting specific regions of the city, specifying date ranges, zoom in/out capabilities, and shows the total number of active cases, recovered cases, and deaths.  

\noindent \textbf{Concluding Remarks and Ongoing Efforts:} We conclude our introduction with some final remarks and outline our current research efforts.
 
\begin{enumerate}[leftmargin=*]

\item The large scale nature of our study where we discern and model the mobility patterns of 2 million users each day in Rio de Janeiro and its suburbs, the  second most populous city in Brazil, coupled with the fact that most countries around the world have failed to effectively contain the pandemic provides us the footing to confidently hypothesize that  the observations and conclusions drawn here can be extrapolated to cities around the world.   

\item Overall, our research reveals that while lockdowns reduced the amount of human mobility, a high (approximately 15\%) of the population still ventured significantly out of their neighborhood, which could have partially contributed to our failure in containing the spread of COVID-19.  With COVID-19 cases once again on the rise and countries around the world bracing for a second wave, our analysis shows that if governments resort to lockdowns as a measure to contain the disease, stricter implementation of lockdown measures may be necessary to decrease the mobility of people.  

\item Our analysis and the interactive website arms government authorities with scientific analysis and  tools to design and implement effective policies to contain the current pandemic.  Importantly, the learnings from this work along with our ongoing research on designing and integrating mobility prediction models can enable authorities to take minimally invasive actions (e.g., traffic rerouting, city planning) to avert a surge in infections in place of widely unpopular blanket lockdown interventions \cite{youtube_fauci}. Additionally, as part of our ongoing efforts, we are investigating the correlation between mobility and the number of reported COVID-19 infections, which can further enable us to mitigate the spread.
\end{enumerate}

\vspace{-2mm}
\section{Related Work}
\label{sec:related}

Our work on characterizing mobility during COVID-19 touches upon different areas such as Internet and web measurement studies, mobility analysis and modeling, and analysis of mobile use with ties to geographic locations.  In contrast to existing work, the primary goal of this work is to understand and model human mobility by leveraging cellular data connections during COVID-19, and lays the foundation  for designing analytics-based tools and models to improve societal and governmental preparedness and response. 

Due to the recent nature of the pandemic, there is limited work on measurement studies examining the impact of pandemic on different network parameters. Lutu et al. \cite{lutu2020characterization} characterize the impact of COVID-19 on mobile network operator traffic and analyze the changes brought upon by the pandemic. Feldman et al. \cite{feldmann2020lockdown} analyze Internet traffic during COVID-19 and find that the overall traffic volume increases by 15-20\% within a week of the pandemic. There is also work on measuring the reaction to the pandemic on the Internet and social media \cite{boettger2020,chen2020tracking}. Zakaria et al. \cite{zakaria2020analyzing} analyze the impact of COVID-19 control policies on campus occupancy and mobility via passive WiFi sensing. Trivedi et al. \cite{trivedi2020wifitrace} use passive WiFi sensing for network-based contact tracing for infectious diseases, particularly focused on the COVID-19 pandemic.

Modeling human mobility using cellular network and mobile application data has been a problem that has garnered interest in the last decade. Some notable ones here are predicting human mobility using attentive recurrent neural networks \cite{feng2018deepmove} and spatio-temporal modeling and prediction using deep neural networks \cite{wang2017spatiotemporal}, learning to transfer mobility between cities \cite{10.1145/3366423.3380210}, leveraging cellular network data for understanding fine-grained mobility \cite{fang2020modeling}. Zhang et al. \cite{zhang2017real} develop a real-time model for human mobility using multi-view learning and Zhu et al. \cite{zhu2020spherical} develop spherical hidden Markov model for understanding human mobility. There is also work on analyzing data in relationship to the geographic locations \cite{singh2019urban,10.1145/3366423.3380141}. Cao et al. \cite{cao2017human} conduct measurement studies on the predictability of human movement in a college campus using WLAN measurements. Pattern mining approaches to detect underlying mobility patterns have also been developed \cite{lian2018joint,comito2018mining}. 

Chaganti et al. \cite{chaganti2018cross} Sadri et al. \cite{sadri2018will} develop a continuous model to predict user mobility in a day. Nikhat et al. \cite{nikhat2018analysis} present an analysis of user mobility in cellular networks. There is also work on early detection of gathering events by understanding the traffic flow \cite{zhou2016traffic}. There is also work on measurement studies in networks understanding how users transition across networks \cite{yang2015measurement}, measuring city-wide signal strength \cite{alimpertis2019city}, modeling mobility using a mixed queueing network model \cite{chen2012mixed}, empirical characterization of mobility of multi-device Internet users \cite{trivedi2020empirical}, and quantitatively evaluating different mobility approaches across different architectures \cite{chaganti2018cross}. There is also extensive work indoor localization and location prediction \cite{margolies2017can,wang2017csi,wang2017csi,AlyHapi}, which are also related to human mobility prediction and analysis.

\section{Data and Methods}
\label{sec:data}

In this section, we describe the cellular network traffic datasets that we collect and use in our analysis. We collaborate with one of the main cellular network providers in Brazil, TIM Brazil, and collect and analyze cellular network connections from all users using this cellular provider in the city of Rio de Janeiro and its suburbs.

Our goal is to analyze and compare cellular network usage data (comprising of phone calls, 3G/4G data connections, and text messages) from  pre-lockdown, during lockdown, and post-lockdown phases  to  understand  and model human mobility patterns during the COVID-19 pandemic, and evaluate the impact of lockdowns on mobility. We collect and log individual cellular connections made by users to approximately 1400 cellular antennas in and around the city of Rio de Janeiro and its suburbs during each 5-minute interval from March 1, 2020 to July 2, 2020. The data consists of approximately 120 million connections logs for each day during this time period, which encompasses approximately 2 million users per day. Overall, the entire dataset comprises of approximately 10 billion connection logs. As Brazil enforced strict lockdown measures from the third week of March to the end of May, the data provides valuable information on human behavior and mobility in the pre-lockdown, during lockdown, and post-lockdown time periods. We note that the cellular network provider has anonymized the data to ensure user privacy.
  \begin{table}[ht!]
\caption{Example instances from the aggregate dataset}
\vspace{-2 mm}
\begin{center}
\noindent\setlength\tabcolsep{2pt}%
\begin{tabular}{|c|c|c|c|}
\hline
\textbf{Timestamp} & \textbf{Latitude} & \textbf{Longitude} & \textbf{Connections}\\
\hline
26th April, 2020, 00:00:00&-20.837028&-43.563111&262\\\hline
timestamp-2&-22.269889&-42.798028&5\\\hline
\end{tabular}
\label{exampleTotal}
\end{center}
\vspace{-3mm}
\end{table}
\begin{table}[ht!]
\caption{Example instances from the individual dataset}
\vspace{-2 mm}
\begin{center}
\noindent\setlength\tabcolsep{4pt}%
\begin{tabular}{|c|c|c|c|}
\hline
\textbf{Timestamp} & \textbf{User ID} &\textbf{Latitude} & \textbf{Longitude}\\
\hline
timestamp-1&hash-1&-23.003431&-43.342206\\\hline
timestamp-2&hash-2&-22.8415&-43.278389\\\hline
\end{tabular}
\label{exampleIndividuals}
\end{center}
\vspace{-3mm}
\end{table}

 We investigate mobile connection data at two different granularities in our analysis: i) aggregated data measured at the antenna level, corresponding to the total number of connections made to each antenna (aggregated), and ii) anonymized individual connections made by each mobile device to the antennas (individual). While the aggregated data is available for the entire duration of the study, the individual data is only available from April 5$^{th}$, 2020. We present some details and statistics about the datasets to understand them before proceeding to a more detailed analysis. In Table \ref{exampleTotal}, we present some example instances from the aggregated data. The data represents the number of  connections made to a specific antenna at a specific instance in time. For example, on the day 04-26-2020 at the time 00.00.00, there are 262 connections to the antenna located at the coordinates [-20.837028, -43.563111]. In Table \ref{exampleIndividuals}, we present some example instances from the individual data. Here, each data instance corresponds to a single anonymized user connecting to an antenna at a specific instance in time.



\begin{figure}[ht!]
	\centering
	\includegraphics[width=0.98\columnwidth]{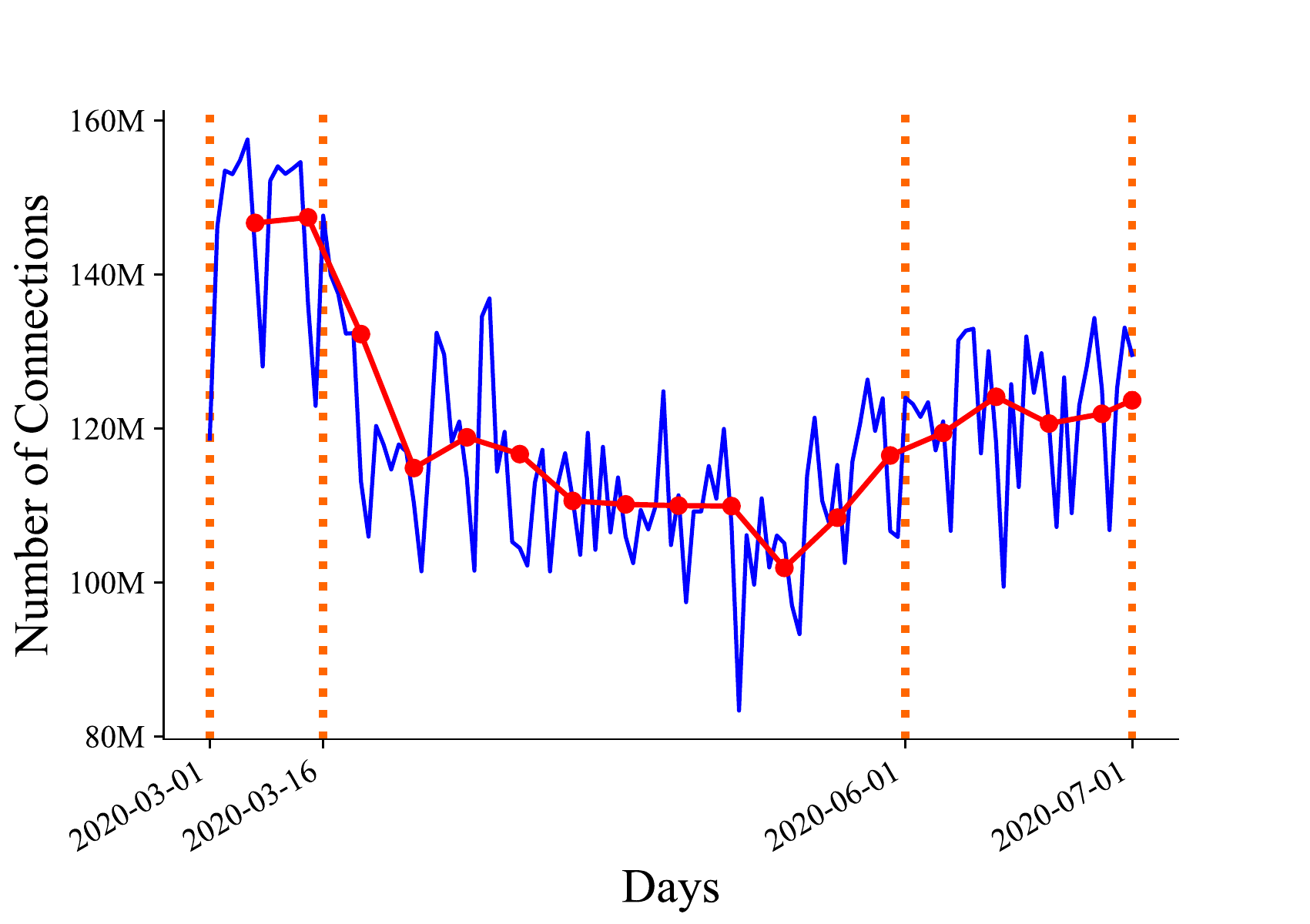}
	\caption{Number of connections per day from March 1, 2020 to July 1, 2020}
	\label{fig:combinedDaysWeeks} 
\end{figure}

\begin{figure}[ht!]
	\centering
	\vspace{-5 mm}
	\includegraphics[scale=0.55]{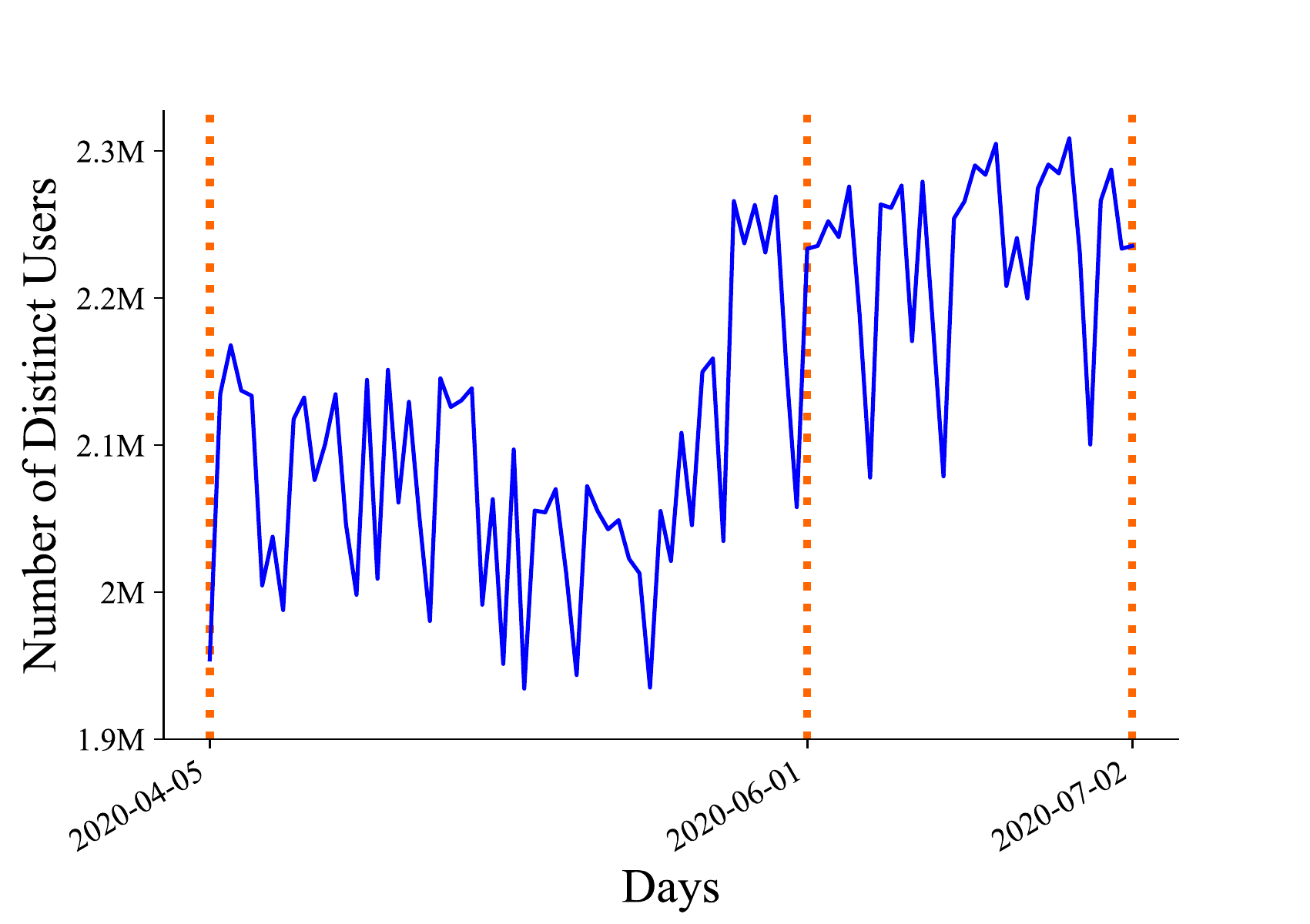}
	\caption{Number of unique users per day between April 5, 2020 and July 2, 2020}
	\label{fig:usersVsDays} 
\end{figure}
Figure \ref{fig:combinedDaysWeeks} shows the total number of connections across the entire duration of our study. Orange vertical lines in the figure separate the different phases in the pandemic: i) the first period from March 1, 2020 to March 16, 2020 represents the period before lockdown, ii) the second period from March 16, 2020 to June 1, 2020 represents the period during lockdown, and iii) the third period from June 1, 2020 to July 1, 2020 represents the period after lockdown when the lockdown measures were eased. The red line captures the average number of connections per week. From the figure, we can see that the total number of connections shows a steadily decreasing trend when lockdown measures are imposed, followed by a period of low connection activity, and then a gradual increase even before lockdown measures are eased. The number of connections continue to increase after the lockdown is lifted, but still the numbers are lower than the pre-lockdown period. 

Figure \ref{fig:usersVsDays} represents the number of distinct users per day from April 5, 2020 to July 2, 2020 in the individual dataset. We again see a reduced number of user connections during the lockdown period when compared to the period after lockdown. From our initial analysis, we observe a discrepancy in the data collection for three days (May 5, 2020, June 21, 2020, and June 22, 2020). We exclude data from these days and then normalize the rest of the week's data for a seven-day period so that it does not interfere our interpretations of change in mobility due to COVID-19.

\section{Socio-Economic and Geographic Background on Rio De Janeiro}
\label{sec:rio}

\begin{figure}[ht!]
	\centering
	\includegraphics[width=0.98\columnwidth]{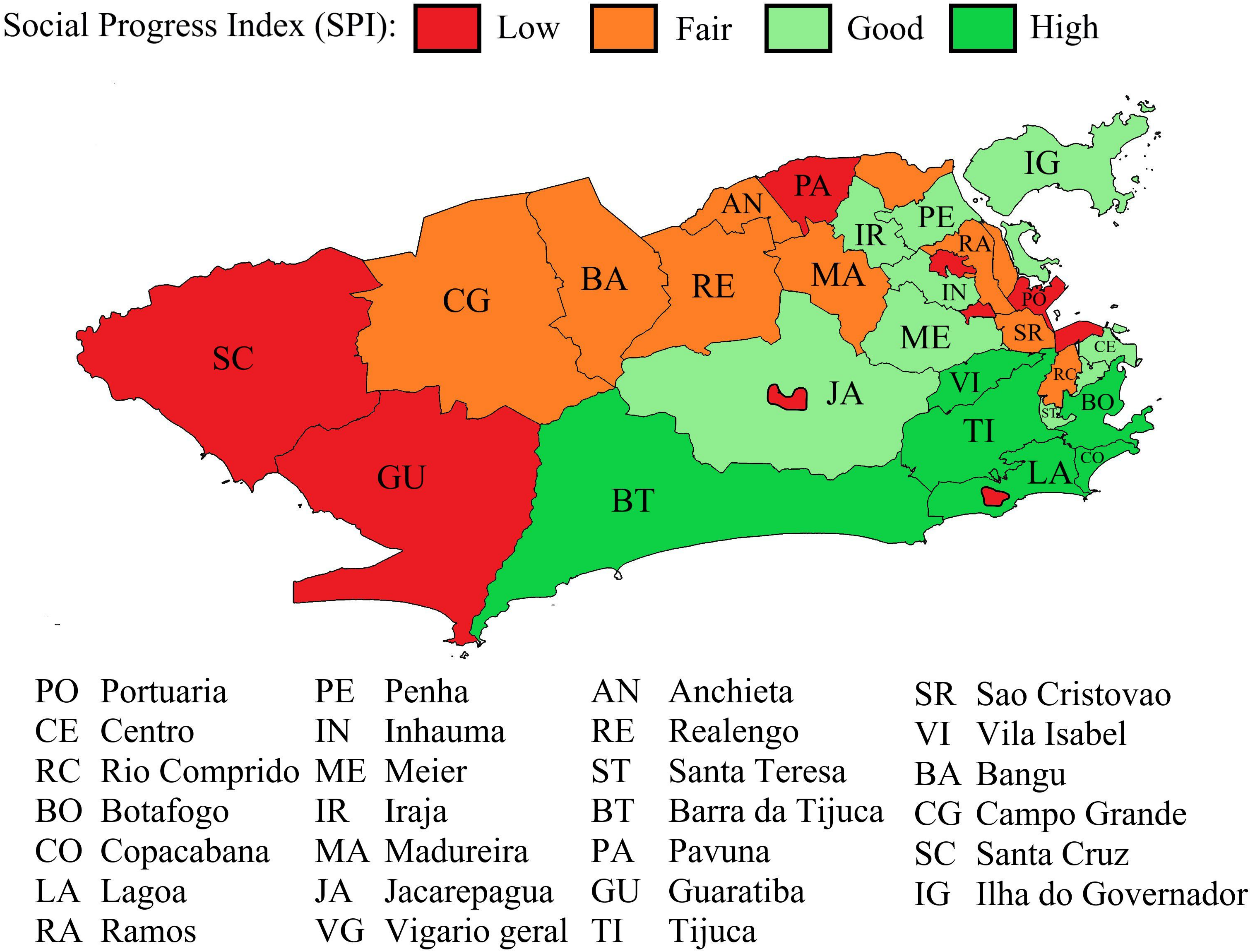}
	\caption{{  The municipality administrative regions in Rio de Janeiro and their respective Social Progress Index (SPI)}}
	\label{fig:regions} 
\end{figure}

To facilitate a better understanding of our analysis and inferences on human mobility patterns, in this section, we provide background on the socio-economic and geographic distribution of people in the different regions of Rio de Janeiro. To this end, we consider the map in Figure \ref{fig:regions}, which depicts the municipality administration categorization of the city and indicates the Social Progress Index (SPI) for each administrative region of Rio de Janeiro. The SPI is a universal performance metric that captures the socio-economic situation \cite{socialprogress}. The map illustrates the administrative regions with low, fair, good, and high SPIs that can be correlated to the average quality of life of the population in these regions. This background fundamentally drives our analysis, and provides us with the necessary context to pose important research questions and draw insightful conclusions.

Rio de Janeiro, like most major cities in the world, has a significant number of companies and business establishments in the downtown area (indicated on the map as CE) or in the surrounding neighborhoods (PO - Portuaria, RC - Rio Comprido, BO - Botafogo, CO - Copacabana). Additionally, while some of the regions (e.g., BO - Botafogo, CO - Copacabana, LA - Lagoa, TI - Tijuca, BT - Barra da Tijuca) are relatively close to the downtown (i.e., a short driving distance away or easy to access via public transportation), some other regions (such as SC - Santa Cruz, CG - Campo Grande, BA - Bangu, RE - Realengo, JA - Jacarepagua) are pretty far and/or very time consuming to commute to/from downtown daily.

Another important detail to note is that people/families with higher socio-economic status usually live near the coast. Administrative regions such as BO - Botafogo, CO - Copacabana, LA - Lagoa, TI - Tijuca, BT - Barra da Tijuca are more expensive and, consequently, have a higher SPI. On the other hand, the lesser privileged population tends to live farther from the coast/downtown. The administrative regions such as SC - Santa Cruz, CG - Campo Grande, BA - Bangu, and RE - Realengo have a lower SPI. One important thing to note is that although Jacarepagua (JA) has a high SPI, there are some parts within this region with a lower SPI. One of the larger lower SPI regions within JA is highlighted on the map (known as Cidade de Deus area).

\section{Connectivity and Mobility Analysis}
\label{sec:experiments}

In this section, we present analysis on the aggregate and individual data to answer the following questions on user connectivity and mobility:
\begin{enumerate}[leftmargin=*]
    \item Which antennas/locations correspond to the maximum user connectivity and traffic and how do they vary during the different phases of the pandemic?
    \item What percentage of users are mobile/stationary and how does that vary during the different phases of the pandemic?
    \item How does user mobility change with the day of the week and how does that vary during the different phases of the pandemic?
    \item What antennas in each region attract the maximum number of users in a week and how do they vary during the different phases of the pandemic?
\end{enumerate}
\subsection{Connectivity Analysis}
\begin{figure}[H]
	\centering
	\includegraphics[scale=0.55]{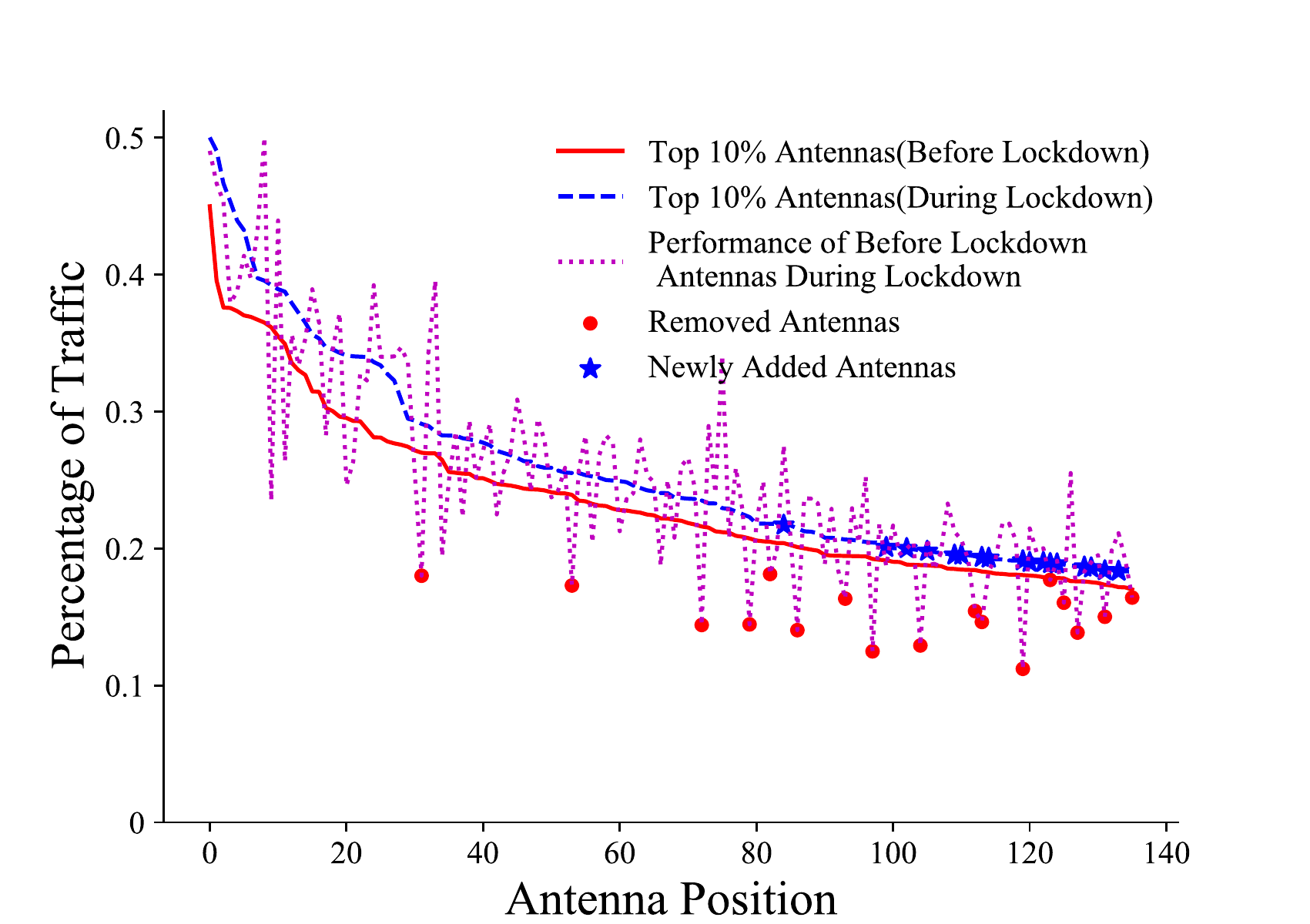}
	\caption{{Percentage of traffic in top 10\% antennas before and during lockdown}}
	\label{fig:BLvsDl} 

	\includegraphics[scale=0.21]{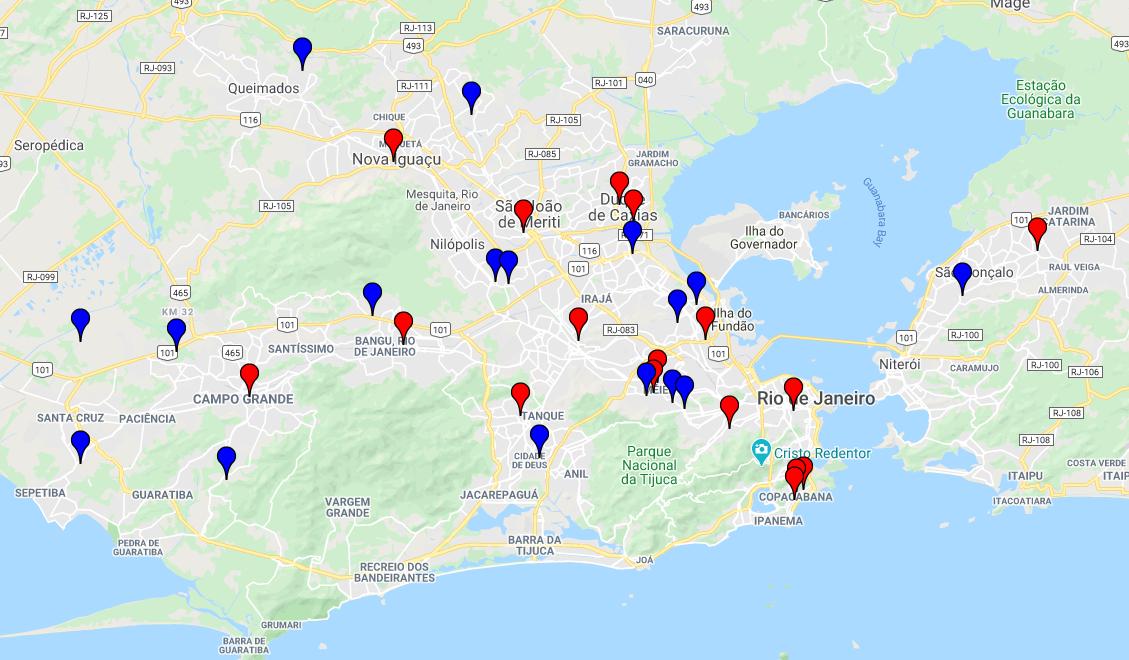}
	\caption{{Differences in the locations of top 10\% antennas before and during lockdown. Blue markers: antennas that newly emerge in the top 10\% during lockdown, Red markers: antennas that were in top 10\% before lockdown but not in top 10\% during lockdown.}}
	\label{fig:mapBLvsDL} 
\end{figure}

We first conduct analysis on aggregate connectivity data and present results on how the connectivity in the top antennas change across pre-lockdown, during lockdown, and post-lockdown time periods. Figure \ref{fig:BLvsDl} shows the percentage of traffic supported by the top 10\% antennas. We first consider the top 10\% antennas before lockdown (solid red line) and the top 10\% of antennas during lockdown (dashed blue line), both ordered in decreasing order by the percentage of traffic supported by them. We observe that the top 10\% antennas during lockdown overall incur a higher percentage of traffic when compared to the top 10\% of antennas before lockdown even though the absolute amount of traffic during lockdown is lower than pre-lockdown (Figure \ref{fig:combinedDaysWeeks}). Now, to understand the difference between the traffic in the antennas before and during lockdown, we consider the same set of antennas that garner the top 10\% traffic before lockdown and plot the traffic supported by them during the lockdown period (dotted magenta). We keep the order of antennas here same as the solid red line (top 10\% before lockdown in decreasing order of traffic) to enable an easier visual comparison. We observe that some antennas incur a significantly lesser percentage of traffic during lockdown (the low points in the dotted magenta line). We also observe that some of these antennas forfeit their position in the top 10\% during lockdown (marked as red dots in dotted magenta) and other antennas take their place in the top 10\% during lockdown (blue stars in the dashed blue line).

\begin{figure}[ht!]
	\centering
	\includegraphics[width = 0.96 \columnwidth]{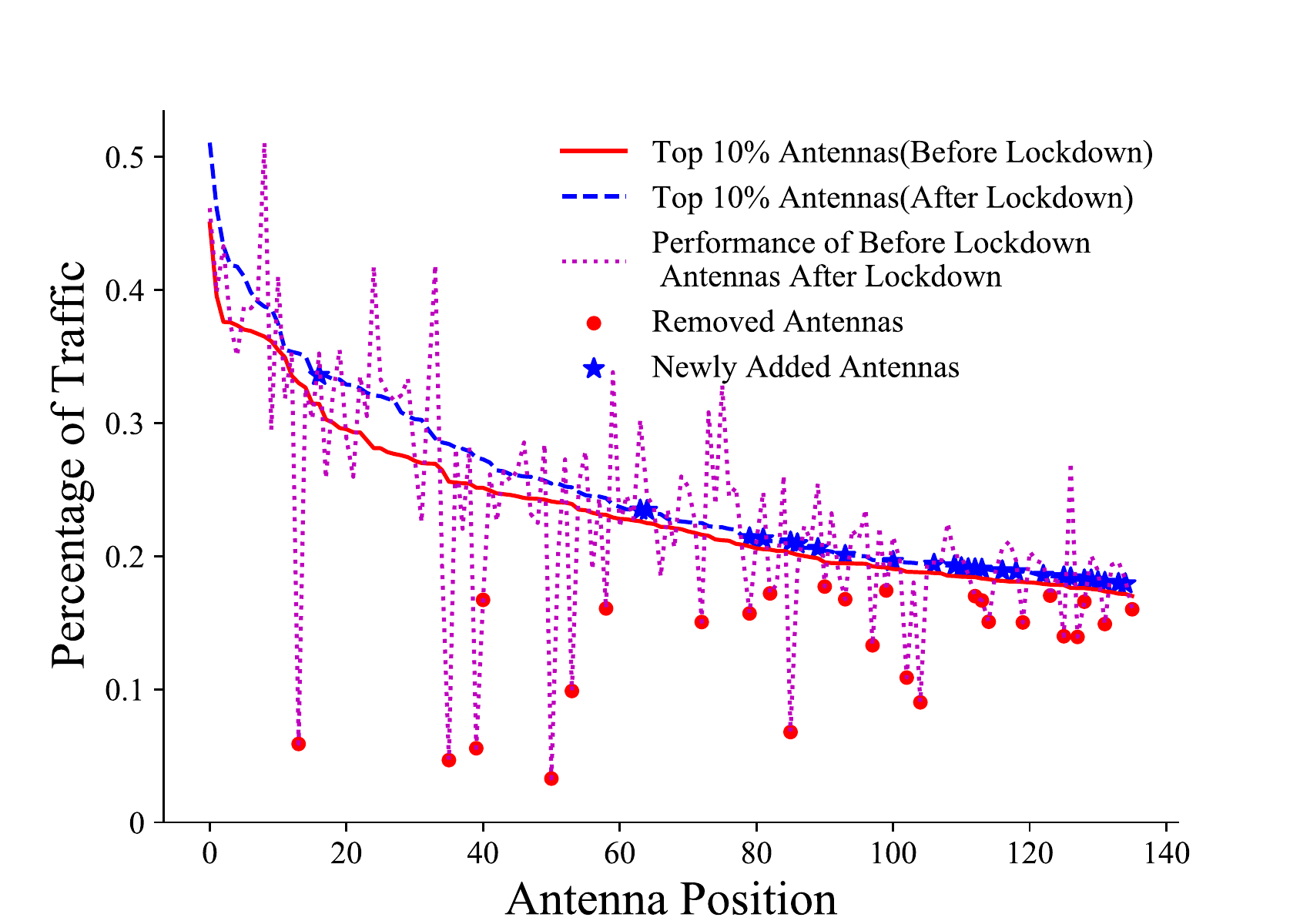}
	\caption{{Percentage of traffic in top 10\% antennas before and after lockdown}}
	\label{fig:BLvsAL} 

	\includegraphics[scale=0.21]{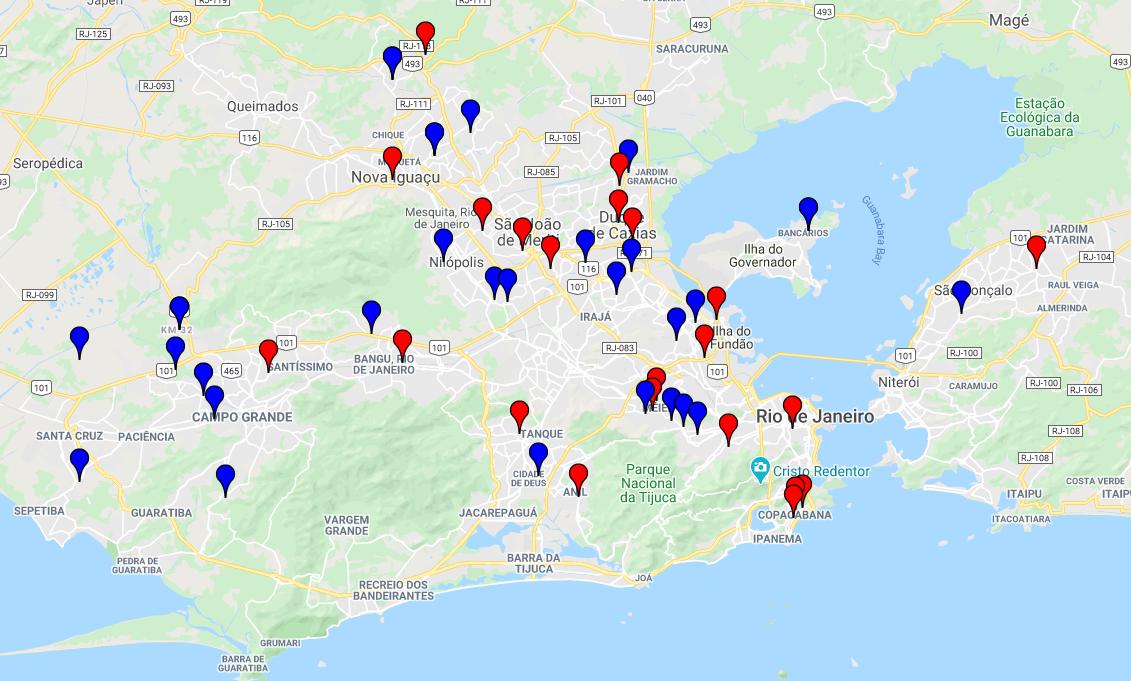}
	\caption{{Differences in the locations of top 10\% antennas before and after lockdown. Blue markers: antennas that newly emerge in the top 10\% after lockdown, Red markers: antennas that were in top 10\% before lockdown but not in top 10\% after lockdown.}}
	\label{fig:mapBLvsAL} 
\end{figure}

Having examined the differences in connectivity in the top 10\% antennas, we proceed to analyze the locations of these displaced antennas. In Figure \ref{fig:mapBLvsDL}, red markers represent the antennas that are in the top 10\% antennas before lockdown but not in the top 10\% of antennas during lockdown (corresponding to the red dots in Figure \ref{fig:BLvsDl}). And blue markers represent the antennas that emerge in the top 10\% during lockdown but are not in the top 10\% antennas before lockdown) in Figure \ref{fig:BLvsDl}. We see that heavily trafficked antennas during lockdown emerge in the suburbs, e.g., SC and GU regions (blue markers in the left side of the map in Figure \ref{fig:BLvsDl}) due to restrictions in mobility due to lockdown. The antennas that get displaced from top 10\% during lockdown are in the downtown regions, e.g., CE and CO in Figure \ref{fig:regions} (corresponding to red markers in the right side of the map in Figure \ref{fig:BLvsDl}). This is expected because majority of people are likely to be working from home during the lockdown period. 


We perform a similar analysis comparing connectivity before and after lockdown (Figure \ref{fig:BLvsAL}). We see a similar trend of the top 10\% antennas after lockdown (dashed blue) incurring heavier traffic than the top 10\% antennas before lockdown (solid red), though the distance between the red and blue lines in Figure \ref{fig:BLvsAL} is smaller than Figure \ref{fig:BLvsDl}. We note that similar to the during lockdown phase, the total traffic post-lockdown is still lower than the pre-lockdown phase (Figure \ref{fig:combinedDaysWeeks}). This shows that even after the lockdown measures are lifted, the traffic pattern still hasn't returned to normalcy. Similar to the comparison between pre-lockdown and during lockdown, we see changes in the antennas that contribute to the top 10\% during the pre-lockdown and post-lockdown periods.  Examining the traffic of the antennas in the top 10\% before lockdown in the post-lockdown time period (dotted magenta), we observe that antennas get displaced from top 10\% from positions as high as the $20^{th}$ percentile (red dots in the dashed magenta) and other antennas take their place (blue stars in the dashed blue line). These graphs serve as motivation to perform a finer grained analysis of connectivity and mobility, which we detail in the following sections.  


Now, analyzing the locations of the displaced antennas in the post-lockdown period, we find that there are a higher number of newly added antennas (blue markers in the map in Figure \ref{fig:BLvsAL}) when compared with the number of new additions to the top 10\% during lockdown (blue markers in the left side of the map in Figure \ref{fig:BLvsDl}). When compared with the lockdown period (Figure \ref{fig:BLvsDl}), we observe newly emerging antennas in the CG region, a densely populated regions with fair SPI. Interestingly, we observe that four main antennas close to the downtown in CE, CO, and BO regions still do not feature in the top 10\% antennas in the post-lockdown period which suggests that even after the lockdown is lifted, majority of people are working remotely and are also  actively avoiding heavily congested areas.

\subsection{User Mobility Analysis}

\begin{figure} [ht!]
     \centering
     \begin{subfigure}[b]{0.48\textwidth}
         \centering
         \includegraphics[scale=0.48]{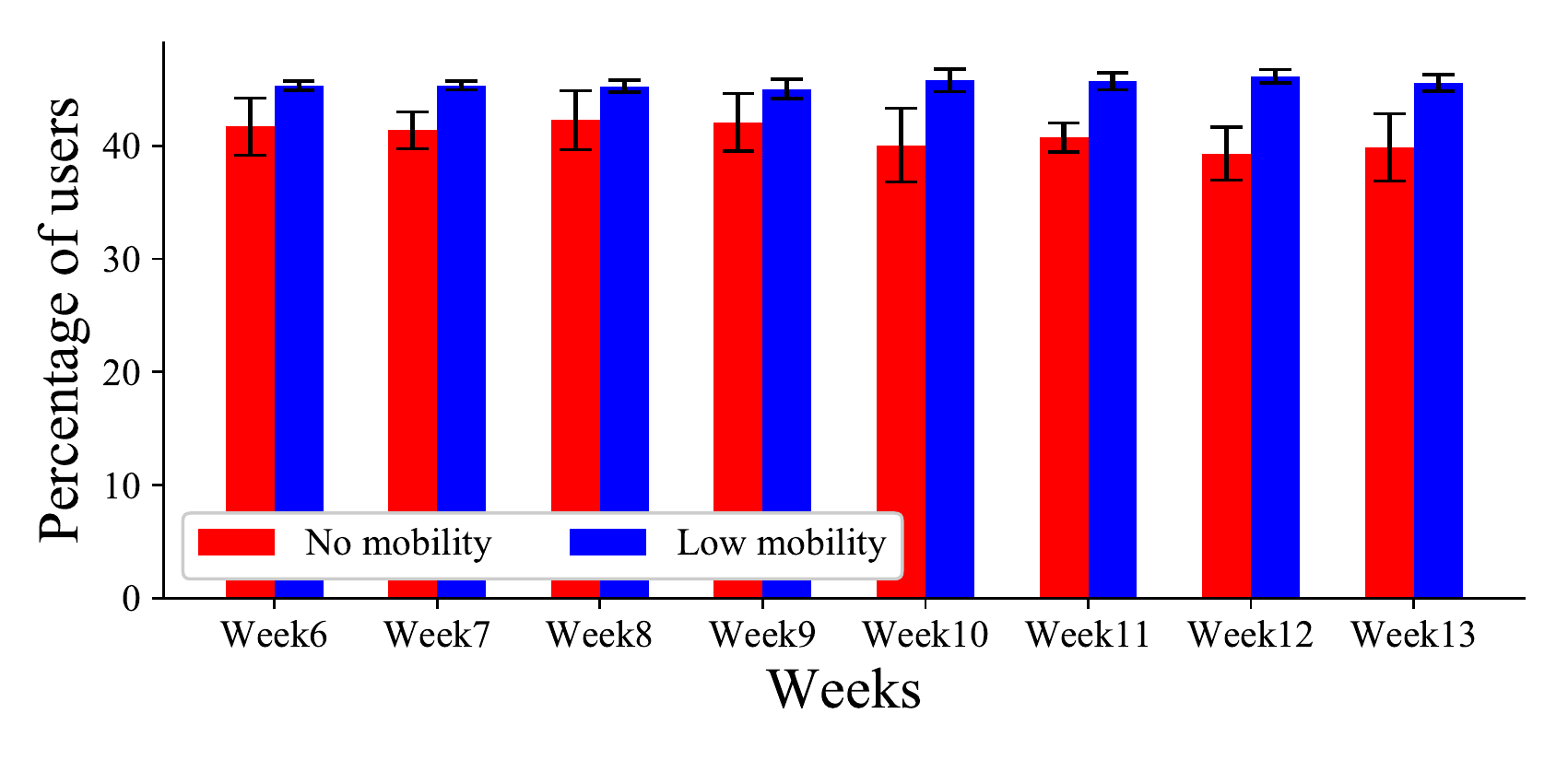}
	\caption{{Percentage of users with low mobility during lockdown}}
	\label{fig:dur1} 
     \end{subfigure}
     \begin{subfigure}[b]{0.48\textwidth}
         \centering
        \includegraphics[scale=0.48]{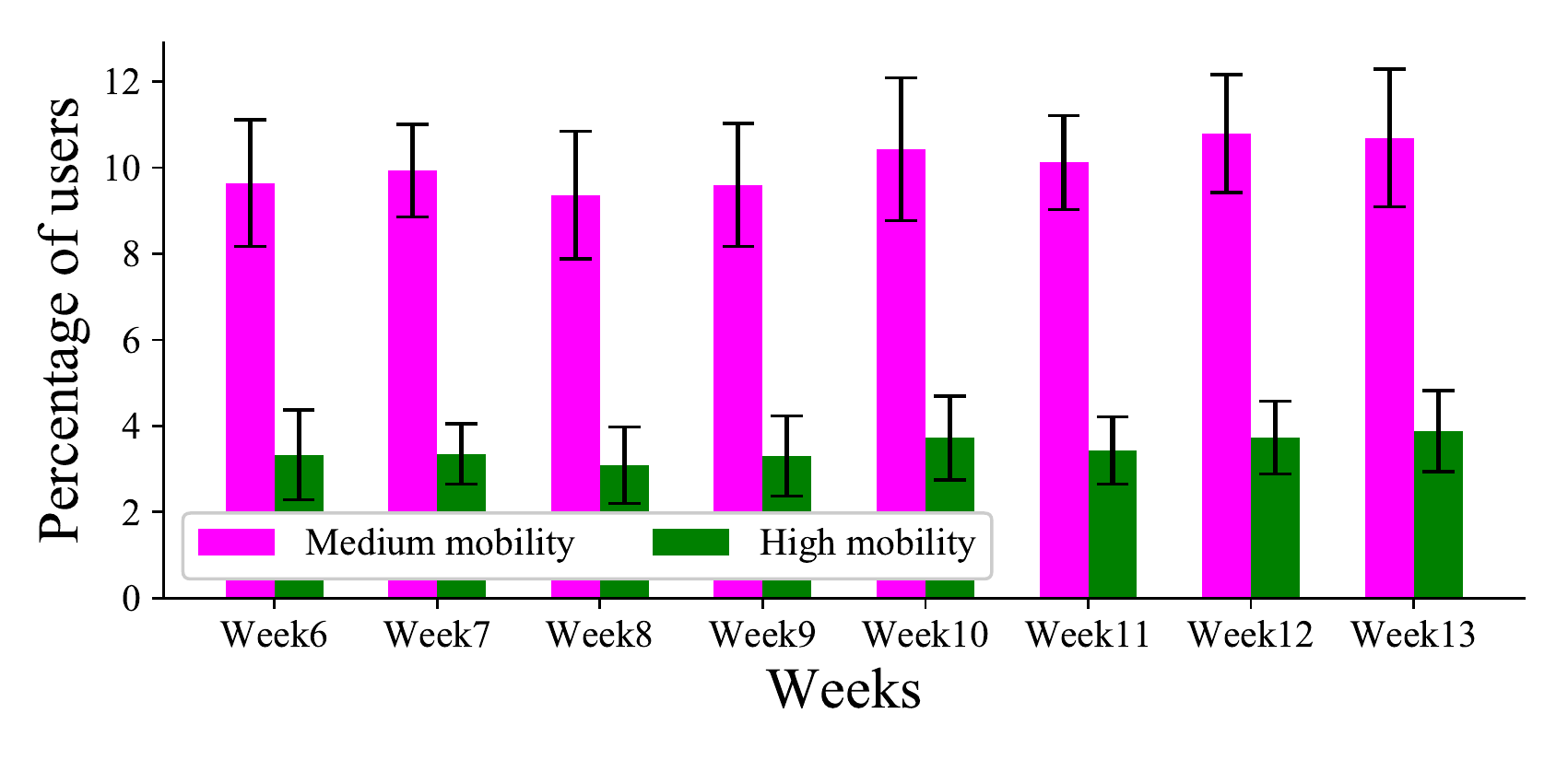}
	\caption{{Percentage of users with moderate to high mobility during lockdown}}
	\label{fig:dur2} 
     \end{subfigure}
     \begin{subfigure}[b]{0.48\textwidth}
         \centering
       	\includegraphics[scale=0.48]{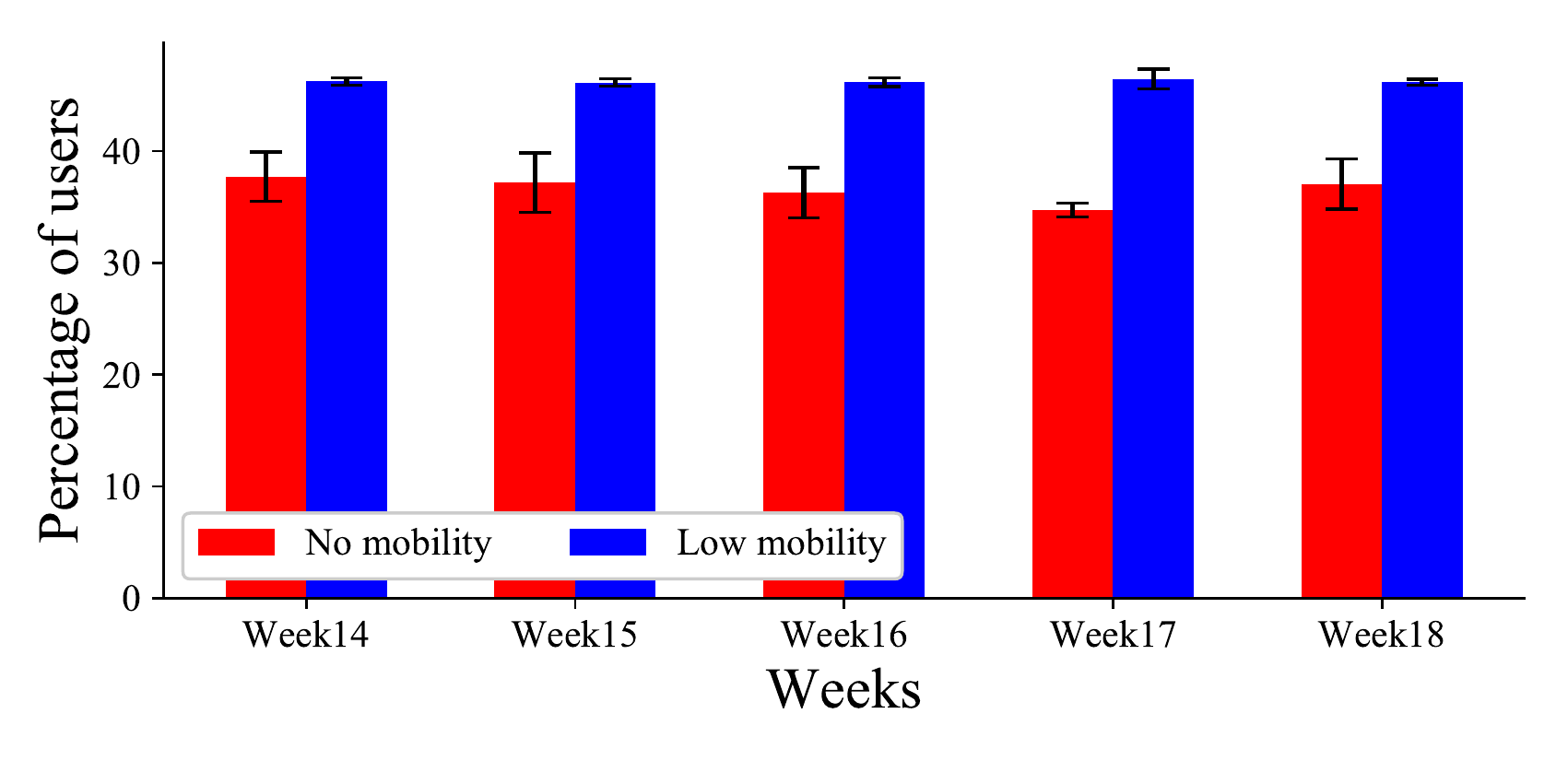}
	\caption{{Percentage of users with low mobility after lockdown}}
	\label{fig:after1} 
     \end{subfigure}
     \begin{subfigure}[b]{0.48\textwidth}
         \centering
	\includegraphics[scale=0.48]{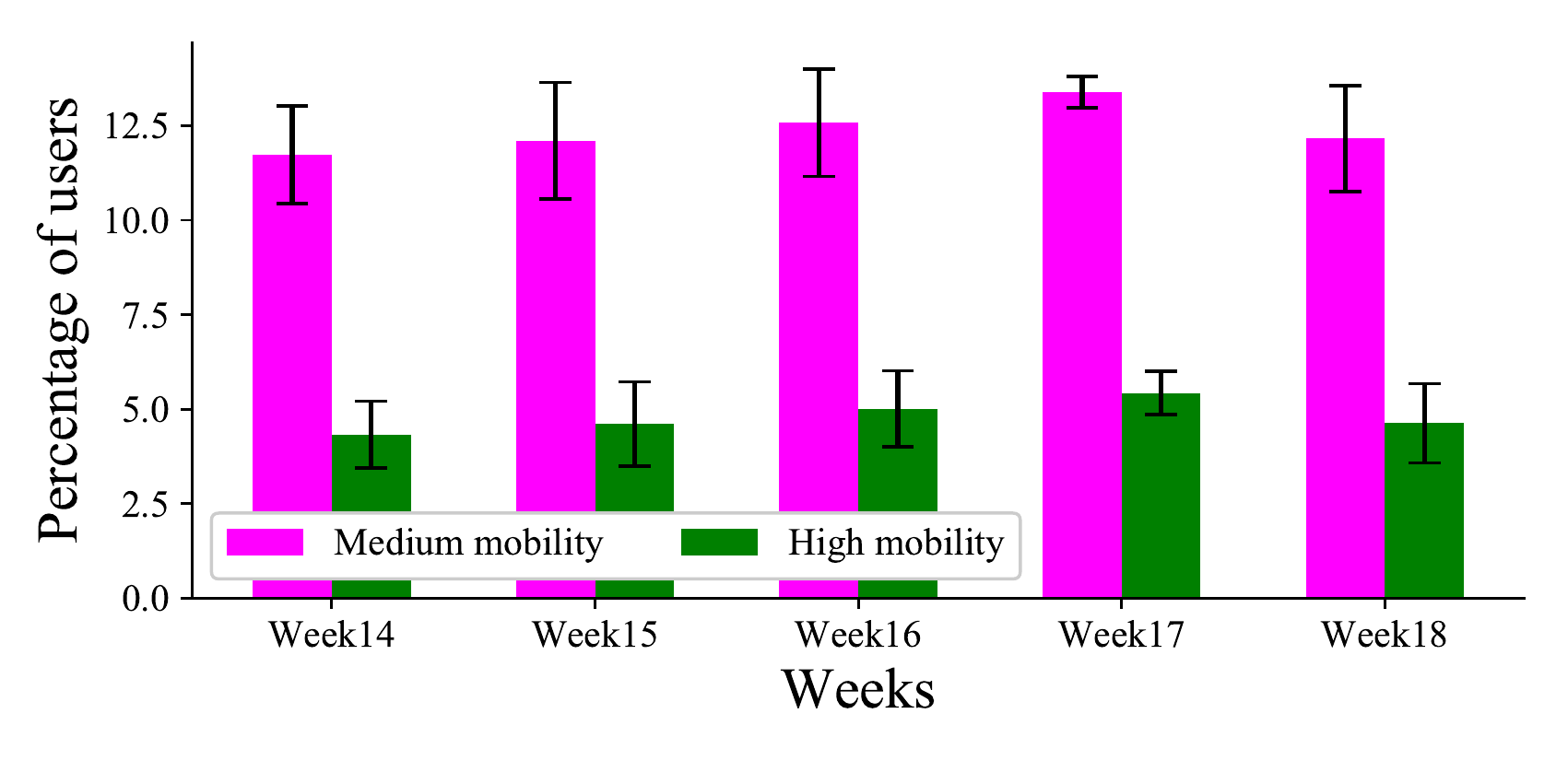}
	\caption{{Percentage of users with moderate to high mobility after lockdown}}
	\label{fig:after2} 
     \end{subfigure}     
     \caption{Comparison of users in the four mobility groups: no, low, medium, and high mobility during and after lockdown}
     \label{fig:threegraphs}
\end{figure}

 Here, we conduct a macroscopic analysis of the mobility of individual users. If a user is only connected to a single antenna, we conclude that the mobility of the user is limited (i.e., the user is primarily indoors; their movement is primarily restricted to the neighborhood where they live). In contrast, if a user is connected to multiple antennas in a day, we conclude that the user is mobile as they must have ventured significantly outside their neighborhood. While it is possible that some users live in an area that is serviced by two antennas, we believe such occurrences are usually rare and do not significantly alter our findings. Thus, the number of antennas a particular user connects to in a day provides a succinct picture of the mobility of the user. We use the individual data for this analysis. Since this data is only available from April 5$^{th}$, our graphs start from week 6 to synchronize the duration of our analysis with the aggregated data.
 
 After calculating the distinct number of antennas for each user per day, we group users based on this number as exhibiting no mobility (i.e., 1 antenna visited), low mobility (i.e.,  between 2 and 5 antennas visited), medium mobility (i.e.,  between 6 and 10 antennas visited), and high mobility ((i.e., greater than 10 antennas visited). Figure \ref{fig:dur1} shows the percentage of users who are exhibiting no/low mobility during the lockdown period. We observe that the number of users with no mobility follows a decreasing trend during the lockdown period from $\sim$40\% to $\sim$35\%, which suggests that more people are venturing out of their homes as the lockdown progresses.  Correspondingly, we observe an increase in the percentage of users with medium mobility as the lockdown progresses (Figure \ref{fig:dur2}). We observe that this trend continues after lockdown as well; the percentage of users in the no mobility group decreases, while the percentage of users in the higher (medium and high) mobility groups continues to increase (Figures \ref{fig:after1} and \ref{fig:after2}). Following the medium mobility users during and after lockdown, we observe an increase from < 10\% at the start of lockdown in week 6 to approximately 13\% in week 17. While the percentage of users in the high mobility group almost remains constant overall during lockdown, we can see a more pronounced increase after lockdown (Figures \ref{fig:dur2} and \ref{fig:after2}). 
 
 The number of users with low mobility remains approximately the same throughout the lockdown period and after the lockdown (Figures \ref{fig:dur1} and \ref{fig:after1}). One possible reason could be that users from the no mobility group may have transitioned to the low mobility group (no mobility users decrease from 40\% to 35\%), while a similar percentage of users transition from the low mobility to medium/high mobility groups. Some users in this group may have also  ``adjusted'' to the new normal and  adapted their mobility patterns around the pandemic for the duration of our study to keep the total percentage approximately the same. In contrast, our analysis in Figures \ref{fig:dur2} and \ref{fig:after2} reveals that approximately 4\%  of users (i.e.., 80K users) visited 10 or more antennas per day, which suggests high mobility for certain individuals. This high mobility can be attributed to essential workers (e.g., sanitation workers, postal workers, taxi drivers) as well as low-income workers who need to travel far for work to sustain their livelihood and families during these trying times. Additionally, the high mobility could also be attributed to individuals who demonstrate less adherence to lockdown rules. From our analysis, we conclude that while the lockdown reduced the amount of human mobility, a high (approximately 15\%) of the population still ventured significantly out of their neighborhood, which could have partially contributed to our failure in containing the spread of COVID-19.


\subsection{Graph-based Mobility Analysis}

Our analysis in the previous subsection demonstrates that a significant number of individuals move across antennas that indicates high mobility in and around the city.  Therefore, to better understand the mobility, we perform a graph-based mobility analysis. We construct a graph where the nodes/vertices correspond to the antennas (i.e., the graph has approximately 1400 vertices). We parse the individual user data and every time a user switches from one antenna to another antenna (referred to as a mobility event), we increase the weight of the edge between those two vertices by one. The so constructed mobility graph thus transforms user connections to mobility events and presents the opportunity to investigate the overall mobility in Rio and its suburbs at an aggregate level.  As we have data loss for May 5, June 21, and June 22, we perform the following pre-processing to ensure fair comparison across weeks. For week 10, we ignore May 5 and scale by a factor of 7/6. For week 17, we ignore June 21 and June 22 and then scale it by 7/5. For week 18, we only have 5 days of data, so we also scale week 18 by 7/5.

\begin{figure}[ht!]
	\centering
	\includegraphics[width=0.97\columnwidth]{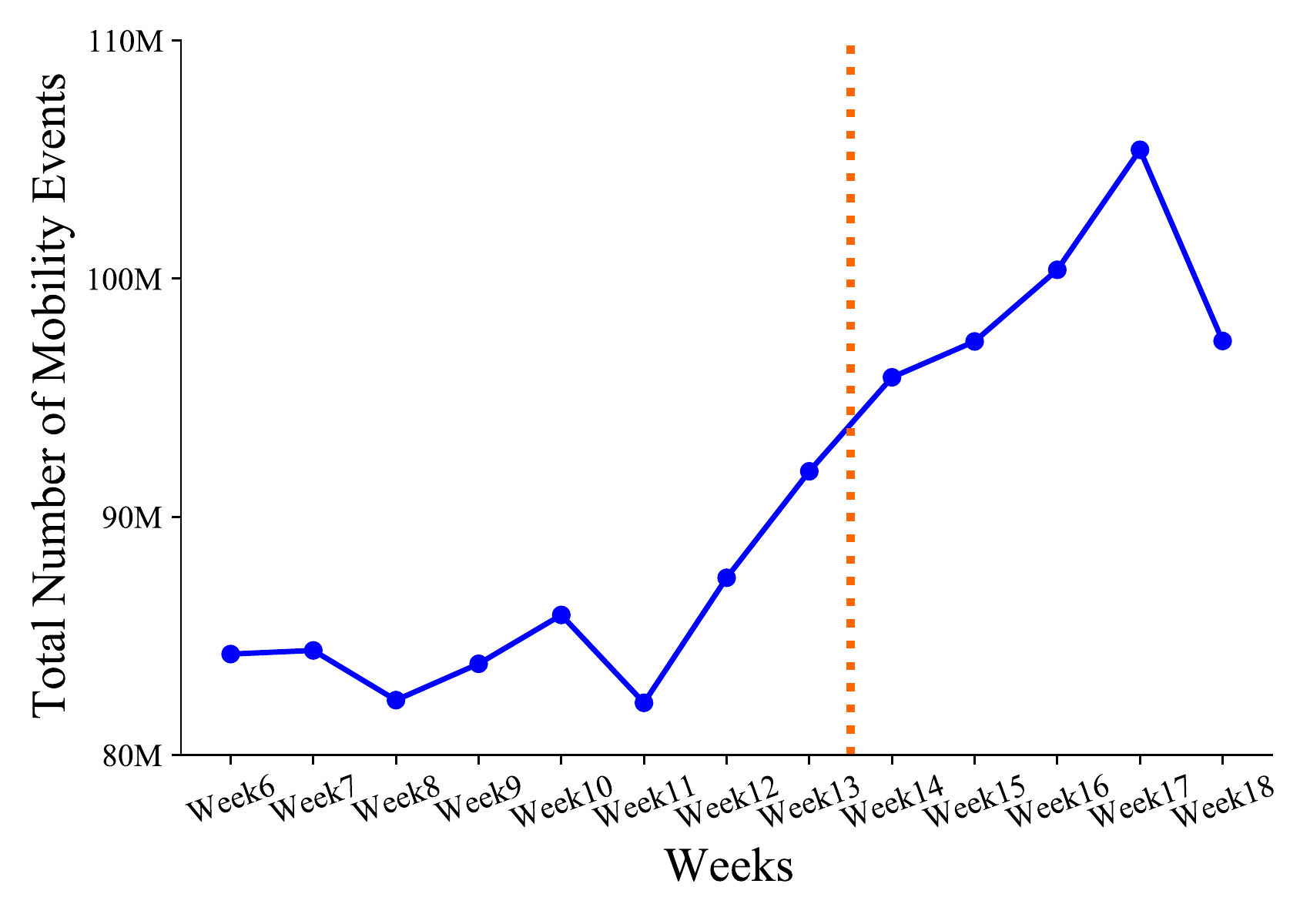}
	\caption{{ Total number of mobility events over weeks}}
	\label{fig:mobilityVsWeeks} 
	\end{figure}
\begin{figure}[ht!]
	\centering
	\includegraphics[scale=0.55]{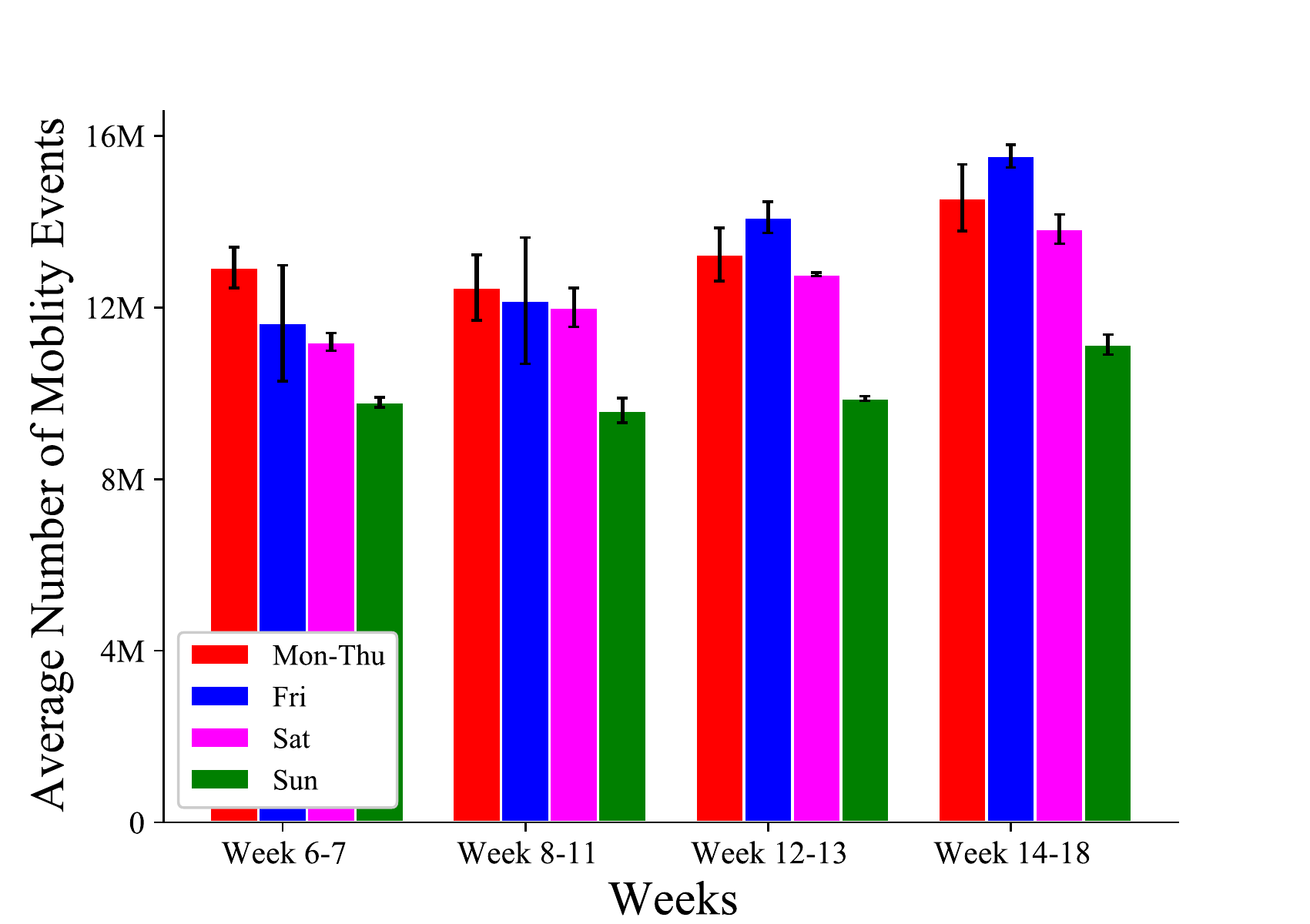}
	\caption{{Average number of mobility events over weeks on different days of the week}}
	\label{fig:mobilityweek} 
\end{figure}

Figure \ref{fig:mobilityVsWeeks} shows the distribution of the total number of mobility events over weeks. The vertical orange line represents the day when lockdown is eased. We observe from the figure that the overall user mobility in the city starts increasing from about three weeks before the lockdown measures are eased. The increase in mobility continues into the post-lockdown period as well. This finding is synchronous with Figures \ref{fig:dur1}, \ref{fig:dur2}, \ref{fig:after1}, and \ref{fig:after2}, which also indicate higher mobility of people with the passage of time. This further establishes that there is a significant increase in mobility when lockdown restrictions are eased. 

\begin{figure}[ht!]
	\centering
	\includegraphics[width=0.97\columnwidth]{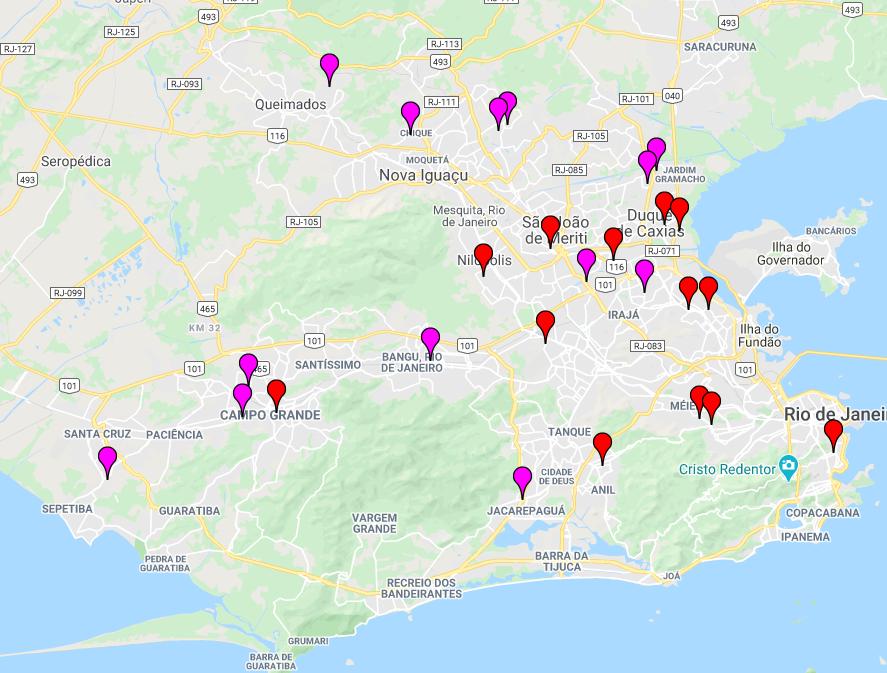}
	\caption{{Differences in locations of top 10\% antennas between weekdays and Saturdays in week 8-11 period. Red markers: antennas in top 10\% during Mon--Thu but not on Saturday, Pink markers: antennas in top 10\% on Saturday that are not in top 10\% Mon--Thu. }}
	\label{fig:mapWeekdaysVsSaturday} 

	\includegraphics[width=0.97\columnwidth]{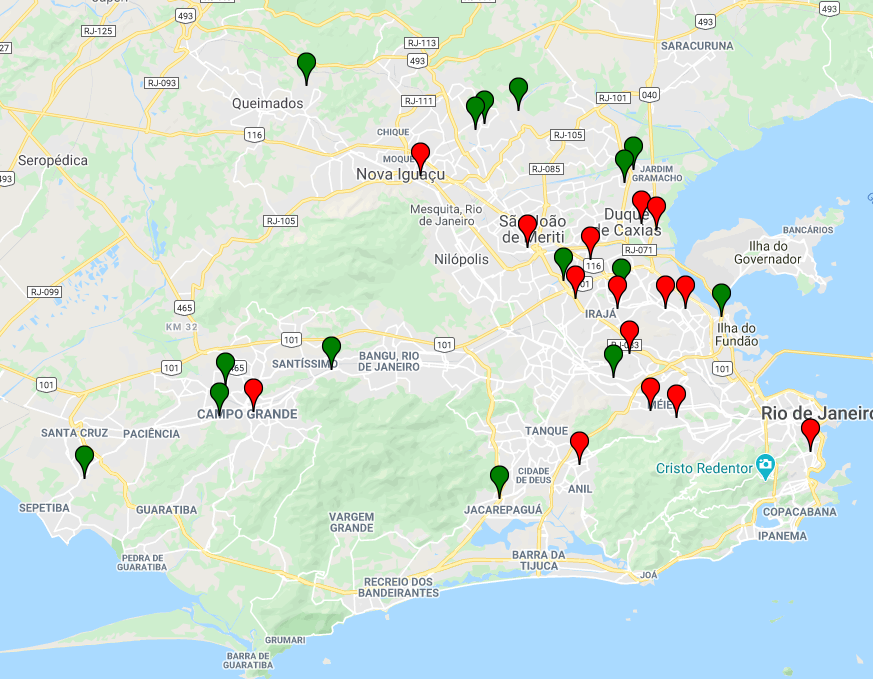}
	\caption{{Differences in locations of top 10\% antennas between weekdays and Sunday in week 8-11 period. Red markers: antennas in top 10\% during Mon--Thu but not on Sunday, Green markers: antennas in top 10\% on Sunday that are not in top 10\% Mon--Thu. }}
	\label{fig:mapWeekdaysVsSunday} 
\end{figure}

We next investigate the impact of the day of the week on the mobility of individuals (Figure \ref{fig:mobilityweek}). To perform this study, we first observe from Figure \ref{fig:mobilityVsWeeks} that the overall mobility pattern remains similar for some weeks. Therefore, we group some weeks together based on their mobility patterns (weeks 6--7, weeks 8--11, weeks 12--13, and weeks 14--18). We then group Monday through Thursday together because they are working days and keep Friday, Saturday, and Sunday as separate days. We consider Friday separately because it captures the mobility pattern of a work day during the earlier part of the day and that of a weekend during the later part of the day. For Mon--Thu, we plot the average of the four days. The error bars capture the variation in the number of mobility events. 

We observe from Figure \ref{fig:mobilityweek} that the overall mobility is lesser on weekends, particularly on Sundays. While one would expect this behavior in a pre-COVID society, we observe that this behavior persists even during lockdown. This additionally suggests that a significant portion of the population still ventures outside for their work during the week and does not have the opportunity to work remotely. Interestingly, we observe that the mobility on Fridays is  lower than  Mon--Thu  during the initial part of the lockdown period and then increases and surpasses Mon--Thu. One plausible explanation is that despite the rising number of cases more individuals are slowly socially self relaxing the lockdown measures and are going to work during the day on Friday and then participating in recreational and/or social activities in the evening, which results in a higher number of mobility events on Friday in comparison to Mon--Thu.

\begin{figure*}
	\centering
	\includegraphics[width=0.98\textwidth]{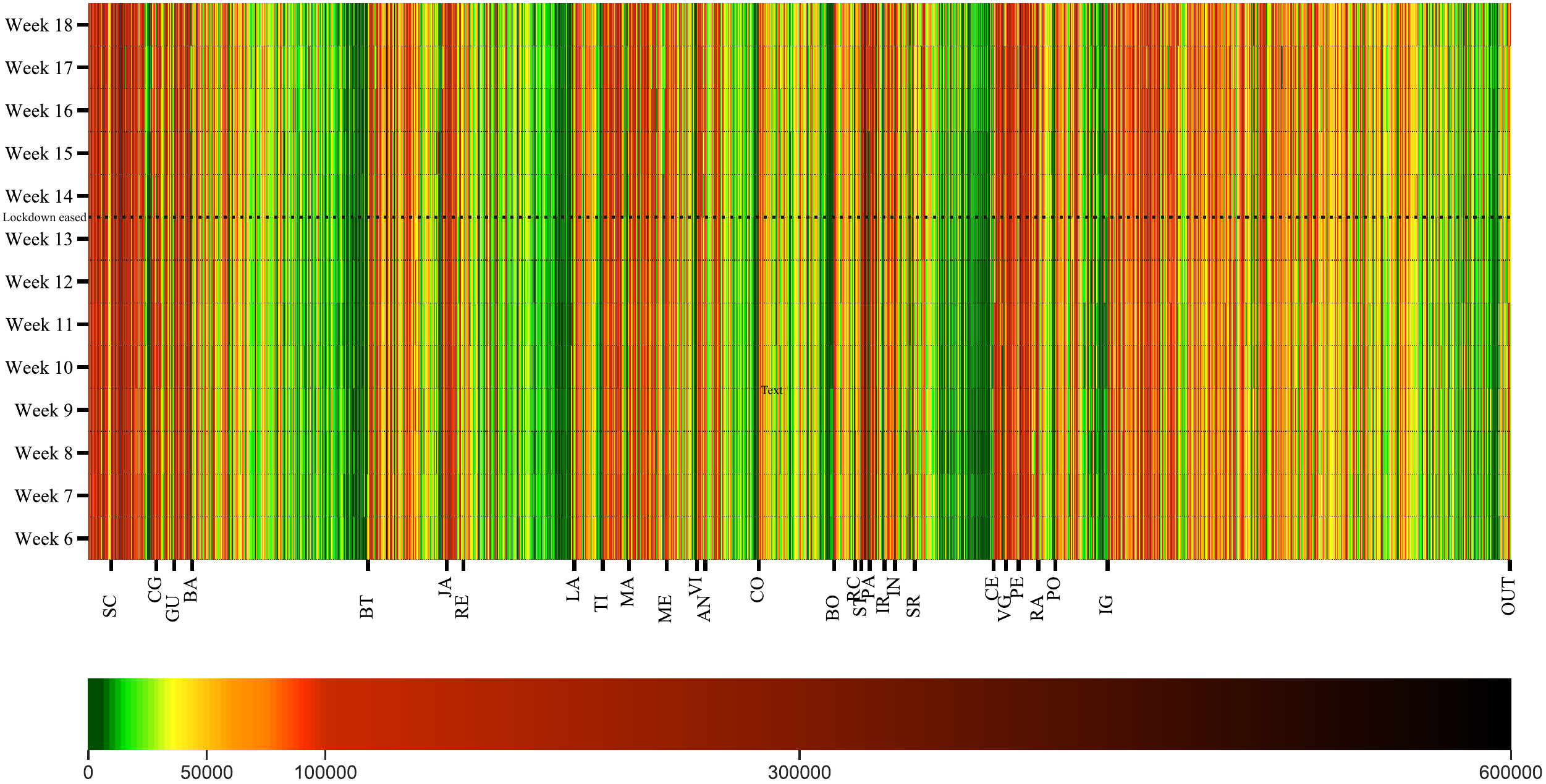}
	\caption{{ Heat map of mobility events for 1400 antennas grouped into regions for during and after lockdown periods}}
	\label{fig:HeatMap_main} 
\end{figure*}
\begin{figure}[ht!]
	\centering
	\includegraphics[scale=0.55]{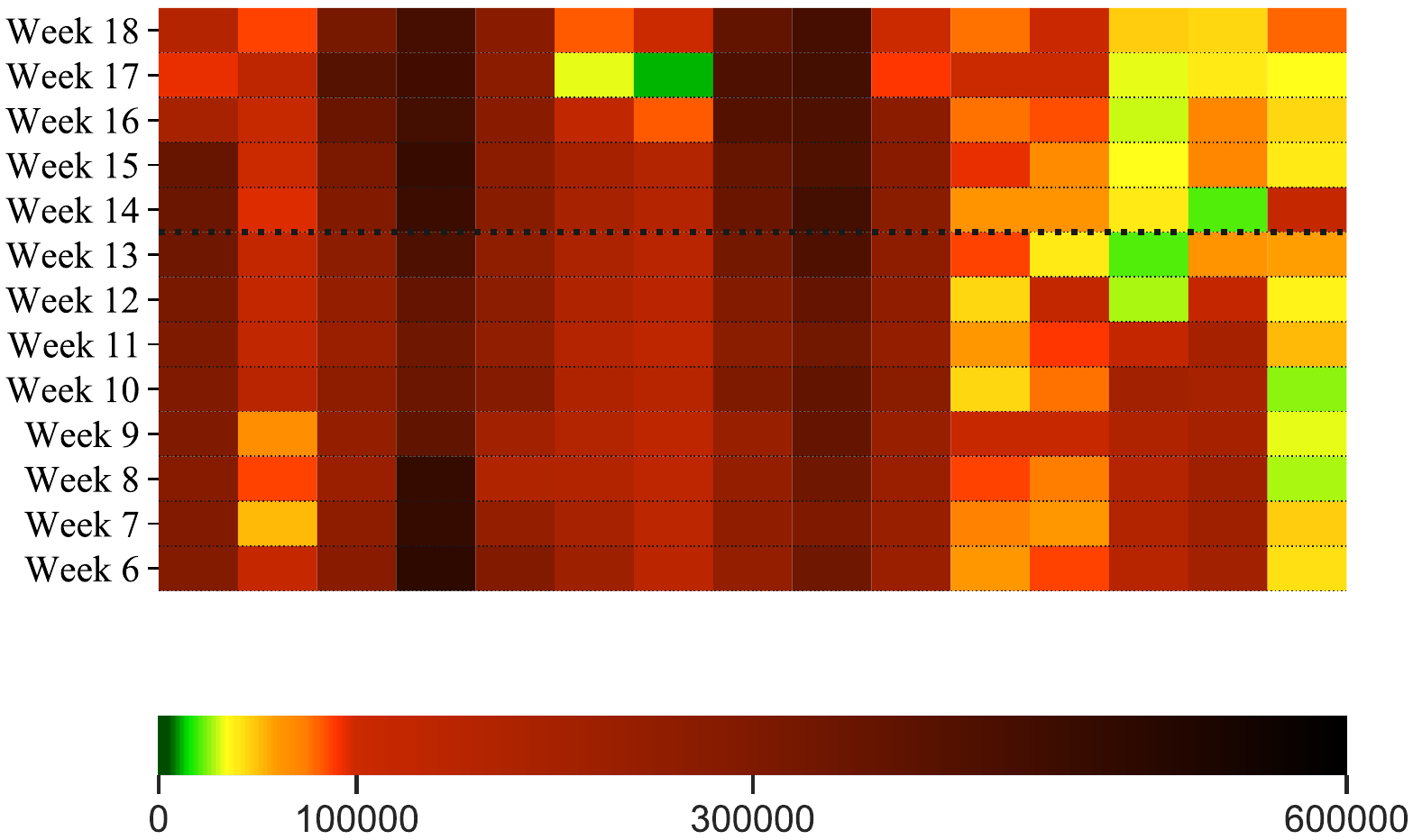}
	\caption{{ Heat map for 15 antennas with the highest variation in the high mobility group (average number of mobility events > 50,000)}}
	\label{fig:heatMap_Greater} 
\end{figure}

\begin{figure}[ht!]
	\centering
	\includegraphics[scale=0.55]{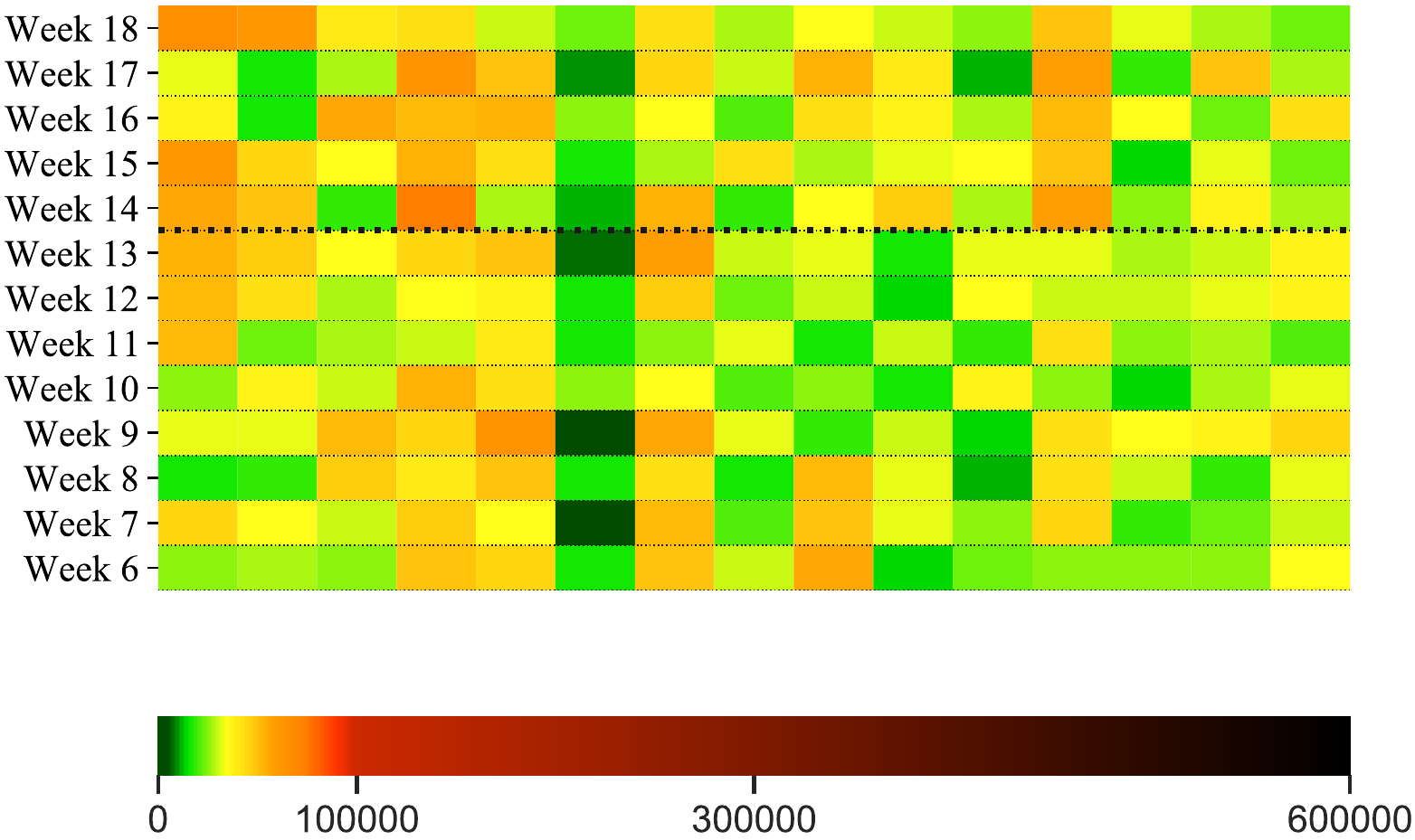}
	\caption{{ Heat map for 15 antennas with the highest variation in the low mobility group (average number of mobility events < 50,000)}}
	\label{fig:heatMap_Less} 
\end{figure}

We next investigate the top 10\% antennas to identify the differences in the connectivity pattern depending on the day of the week. As Friday is a work day, we observe that there is limited variation in the top 10\% antennas, with a difference of 5--8 antennas when compared with the top 10\% antennas in Mon--Thu. In comparison, we observe a significantly higher variation in the top 10\% antennas over the weekend when compared to the weekdays with the difference being higher for Sunday in comparison to Saturday. Figures \ref{fig:mapWeekdaysVsSaturday} and \ref{fig:mapWeekdaysVsSunday} show the variations in the top 10\% antennas for Saturday and Sunday for weeks 8--11, respectively.  The red markers on the map correspond to the ones that were in the top 10\% during Mon--Thu but were replaced with the new antennas shown in pink and green for Saturday and Sunday, respectively. We see that the antennas being replaced from the Mon--Thu group are located in downtown Rio and Duque de Caxias. As expected the new antennas in the top 10\% lie in the more residential areas. There are also similarities in the new locations that emerge in the top 10\% on Saturday and Sunday, suggesting that the locations that gather the top 10\% traffic tend to be similar over the weekend, though the amount of activity is higher on Saturday. 

To better understand the variation in mobility for the 1400 antennas over the weeks, we first group the antennas according to the municipality classification in Rio as shown in Figure \ref{fig:regions}. Figure \ref{fig:HeatMap_main} shows the heatmap outlining the variation in the mobility at each antenna in the various regions for weeks 6 through 18. The vertical ticks on the horizontal axis mark the boundary of each region and coincide with the last antenna located in that particular region. The region OUT signifies all the antennas located in the outskirts or suburban areas of Rio in our dataset. Therefore, the antennas in OUT may not be geographically proximal to one another.

We observe from Figure \ref{fig:HeatMap_main} that the amount of mobility varies significantly across regions. While regions such as SC and CG show high mobility throughout the lockdown and post-lockdown periods, some regions such as BT and CE show low mobility. The low levels of mobility in the CE region, which is in the downtown area, is congruous to Figures \ref{fig:mapBLvsDL} and \ref{fig:mapBLvsAL}, where we observe that the antennas in the downtown region relinquish their position in the top 10\% of antennas in terms of the total traffic. This is primarily due to commercial businesses and offices being closed due to lockdown and their employees working remotely. For the other regions, revisiting Figure \ref{fig:regions} provides us possible context for explaining this difference in mobility. We observe from Figure \ref{fig:regions} that SC and CG are regions with low SPI, while BT is a region with high SPI. We hypothesize that due to this socio-economic disparity, people in SC and CG may be compelled to venture out of their homes for work and personal reasons during this challenging time, while people in BT may have the opportunity to stay indoors. In comparison, some areas such as JA contain a mix of low and high mobility antennas. Again, this can be explained by the presence of sub-regions in JA that lie on the extremes of the SPI spectrum (low and high in Figure \ref{fig:regions}).

We next investigate the antennas that show the highest variation in mobility during the lockdown and post-lockdown periods. To conduct this study, we split the antennas into two groups---antennas whose average number of mobility events is less than 50,000 and antennas whose average number of mobility events is greater than 50,000. Such categorization helps us separately study antennas with high variation in both the low and high mobility groups, respectively. Figures \ref{fig:heatMap_Greater} and \ref{fig:heatMap_Less} show the heat map for the top 15 antennas with the highest variation in this low mobility and high mobility groups, respectively.  We observe that there are a number of antennas that show significant variation. Such mobility variations could plausibly be attributed to fluctuations in the number of COVID-19 positive cases, government policies as well as transitions to/from remote/on-site work that cause higher number of individuals to gather in a specific geographical region covered by an antenna.





\section{COVID-19 Borescope}
\label{sec:website}


With many cities around the world going into lockdown again, analyzing mobility and correlating it with the number of infections is a key and promising factor for controlling the spread of the virus. Thus, another significant contribution of this work is developing a visual/interactive tool called COVID-19 Borescope that helps  government and municipality administrations better understand the evolution of COVID-19 by analyzing the correlation between people's mobility and the infection data reports in different regions of Rio de Janeiro. This powerful tool is still under development, but is already launched and available to the public\footnote{Accessible at: \url{http://gwrec.cloudnext.rnp.br:57074/}}.

Developing a tool to provide insights from our analysis is challenging because it has to: i) be able to support a huge amount of data, possibly from multiple data sources, ii) be scalable, iii) offer the flexibility to integrate new algorithms and data sources, and (iv) be user friendly and intuitive for end users. The details of our tool are described in the following sections.

\begin{figure*}[ht!]
	\centering
	\includegraphics[width=0.95\textwidth]{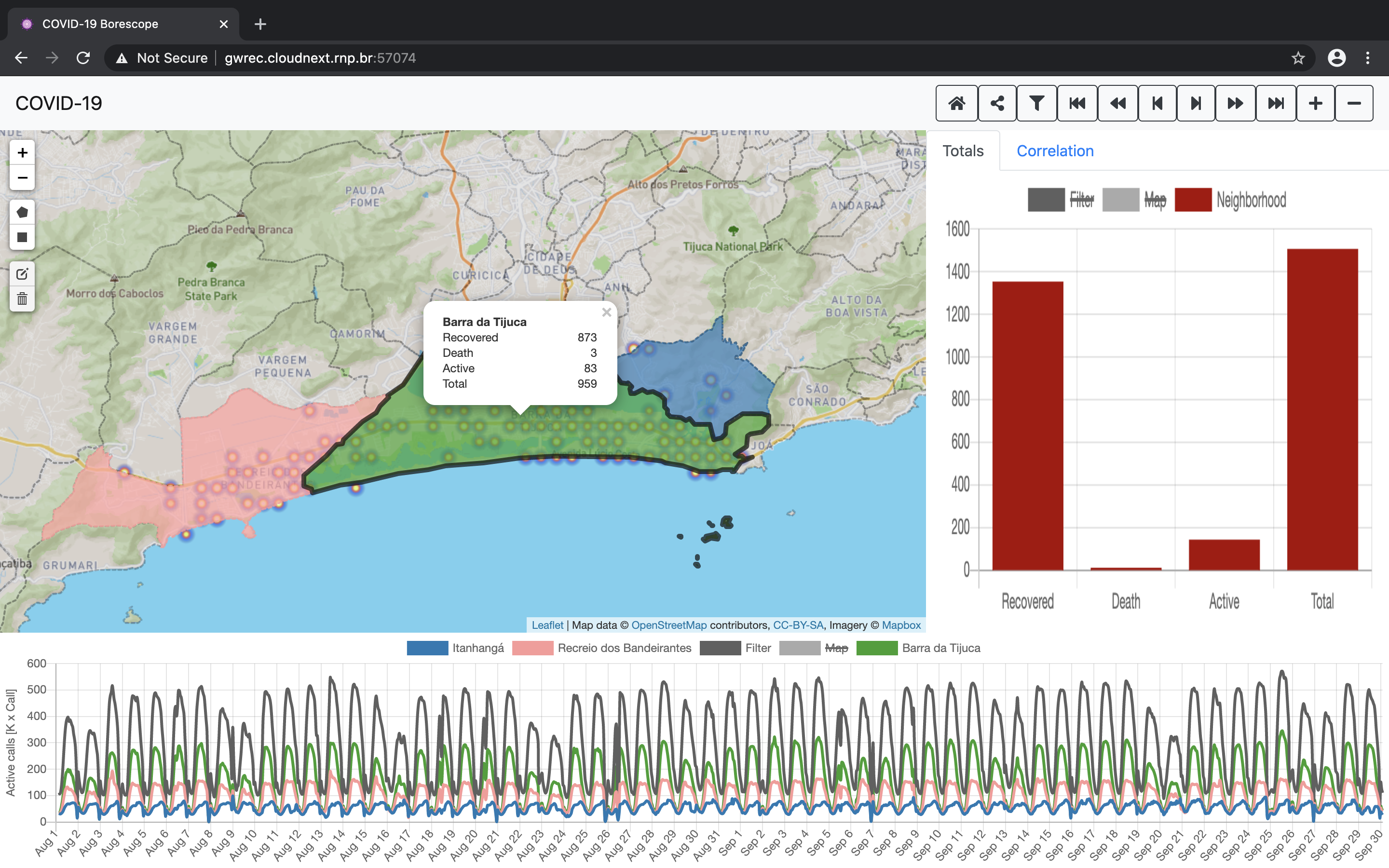}
	\caption{COVID-19 Borescope Web Interface}
	\label{fig:webInterface} 
\end{figure*}

\begin{figure}[ht!]
	\centering
	\includegraphics[scale=0.3]{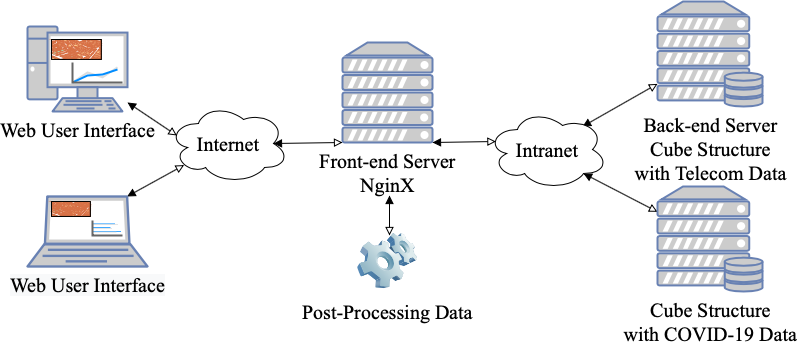}
	\caption{COVID-19 Borescope Architecture}
	\label{fig:Archtecture} 
\end{figure}


\subsection{COVID-19 Borescope Architecture}

To address the challenges involved in the development of a graphical and interactive system that performs intelligent data analysis of visually selected geo-temporal subsets of collected information,  COVID-19 Borescope is supported by a robust underlying architecture. As shown in Figure~\ref{fig:Archtecture}, the architecture consists of three servers, two in the back end and one in the front end.

At the front end, we have NginX\footnote{NginX: \url{http://nginx.org/}} running as the application server. At the back end, we have two data servers, one to store the data received from the cellular network provider and another one to store the data that is obtained from the Open Data repository provided by the Brazilian ministry of health\footnote{Brazilian COVID-19 OpenData:  \url{https://opendatasus.saude.gov.br/dataset}}. The application server processes requests received from the Web User Interface, forwards the processed queries to the appropriate back-end server(s) and waits for the response(s). Once the NginX application server receives the response(s) from the data server, it may either still perform some post-processing or may  directly forward the results to the end user. In the former case, the post-processing scripts are triggered. The tool is very flexible, which makes it possible to include external calls to  machine learning algorithms. In the current version, as part of our ongoing research efforts, we are using this function to analyze correlation between mobility and infection cases in different regions of the city. After finishing the post-processing, the results are sent back to the end user and presented in a graphical manner on the web interface.

\subsection{Core Data Structure}

For the data server, the tool uses a new data structure, which is an optimized variation of  Nanocubes~\cite{nanocubs}. A full description of this data structure is beyond the scope of this paper and will be the subject of a future publication. However, it is important to mention that the data structure is an in-memory database and, as is the case with any Datacube structure~\cite{datacubs}, it is specialized to perform statistical geo-temporal queries in a efficient manner with coordinates organized as QuadTrees~\cite{quadtree}, offering low response time for queries and moderate memory usage.

The data structure uses a JSON-based language that emulates a simplified SQL syntax to retrieve data. It offers the traditional “select”, “where”, and “group by” statements to select, filter, and group/fold data, respectively. The outcome of the data structure is a time series that is forwarded to the end user by the NginX framework, before and/or after being submitted to post-processing functions.

\subsection{Interactive Web Interface}

Figure~\ref{fig:webInterface} provides an overview of the web user interface, which is used for visualization and interactive analysis. We briefly describe the interface. On the top left, the interface provides the option  to select the region(s) in the map the user wants to analyze.  In the bottom of the interface, the graphic shows  the evolution of the number of connections  with time for the selected area of the map. On the right hand side, the "Total" option shows the histogram of COVID-19 reported cases, which includes the number of recovered, active, and total cases, and deaths during the selected time period.  The "Correlation" option shows the result from the correlation analysis.  On the top right, the control panel allows users to zoom in/out, navigate different time periods and filter city regions by name. The pop-up button over the selected neighborhood provides the summary of the COVID-19 numbers for that specific region of the city.

\section{Discussion and Concluding Remarks}
\label{sec:conclusion}

In this paper, we presented a large scale analysis of human mobility during a crucial stage in the COVID-19 pandemic in Rio de Janeiro and its suburbs based on cellular network connection logs from one of the main cellular network providers in Brazil, TIM Brazil. Our analysis employs aggregate and individual data on cellular connections from three phases in the first wave of the COVID-19 pandemic: pre-lockdown, during lockdown, and post lockdown, and draws important conclusions on the impact of lockdown on mobility. Overall, our research revealed that while lockdowns reduced the amount of human mobility, a high (approximately 15\%) of the population still ventured significantly out of their neighborhood, which could have partially contributed to our failure in containing the spread of COVID-19.  Since our analysis is based on large scale data from one of the most populous cities of the world, our analysis and resulting conclusions can potentially have positive implications on understanding mobility and designing lockdowns in other cities in future waves of the COVID-19 pandemic or other future events of a similar nature. With COVID-19 still surging in many countries and cities of the world, we believe our analysis and conclusions can potentially help in the effective management of the pandemic.

Our work opens up avenues for several important research directions. One immediate next step of our study involves studying the correlation between mobility of users and infection rates. A fine-grained understanding of this correlation would be helpful in designing region-specific lockdowns rather than a one-size-fits-all solution, which is challenging to enforce for governments and also hard to adhere for people. Another potential direction involves applying and designing more sophisticated mobility models to understand the patterns more effectively \cite{feng2018deepmove,zhang2017real,zhu2020spherical}. Here, one idea is to study traffic flow patterns to identify bottlenecks and suggest alternate less congested routes and times that can spread the mobility and reduce overcrowding in populous and heavily trafficked areas that have a surge in infection rates. We also plan to develop and integrate mobility prediction models in this effort so that appropriate actions can be taken before a surge in infections occurs. We will continue to integrate our analysis and findings in the COVID-19 Borescope, which presents us with the perfect environment to expand the visibility and utility of our research and potentially help convert that into actionable policies or self-awareness for people.

\bibliographystyle{ACM-Reference-Format}
\bibliography{references}

\end{document}